\documentclass[a4paper,11pt,epsfig,graphicx,subfigure,pstricks]{article}
\usepackage[english]{babel}
\usepackage{epsf}
\usepackage{verbatim}
\usepackage{stmaryrd}
\usepackage{latexsym}
\usepackage{makeidx}
\usepackage{epsfig}
\usepackage{amsfonts}
\usepackage{amsmath}
\usepackage{amssymb}
\usepackage{color}
\input xy
\xyoption{all}
\makeindex
\large \normalsize
\def \epsilon {\varepsilon}
\def\norm#1{\left |#1\right |}
\def\Norm#1{\left \|#1\right \|}
\newcommand{\snot[1]}{{\scriptstyle \times 10}^{#1}}

\title{Semi-analytic computations\\  
of the speed of Arnold diffusion
along single resonances\\ 
in a priori stable Hamiltonian systems}

\author{Massimiliano Guzzo 
\\ \footnotesize Universit\`a degli Studi di Padova 
\\ \footnotesize Dipartimento di Matematica `Tullio Levi-Civita'
\\ \footnotesize Via Trieste 63 - 35121 Padova, Italy
\and 
Christos Efthymiopoulos
\\\footnotesize Research Center for Astronomy 
\\\footnotesize  and Applied Mathematics
\\\footnotesize Academy of Athens
\\\footnotesize Soranou Efessiou 4, 11527 Athens, Greece
\\ \footnotesize 
\and 
Roc\'io I. Paez
\\ \footnotesize Universit\`a degli Studi di Padova 
\\ \footnotesize Dipartimento di Matematica `Tullio Levi-Civita'
\\ \footnotesize Via Trieste 63 - 35121 Padova, Italy
}

\begin{document}

\maketitle

\begin{abstract}
Cornerstone models of Physics, from the semi-classical mechanics in
atomic and molecular physics to planetary systems, are represented by
quasi-integrable Hamiltonian systems. Since Arnold's example, the
long-term diffusion in Hamiltonian systems with more than two degrees
of freedom has been represented as a slow diffusion within the `Arnold
web', an intricate web formed by chaotic trajectories.  With modern
computers it became possible to perform numerical integrations which
reveal this phenomenon for moderately small perturbations. Here we
provide a semi-analytic model which predicts the extremely slow-time
evolution of the action variables along the resonances of multiplicity
one. We base our model on two concepts: (i) By considering a
  (quasi-)stationary phase approach to the analysis of the Nekhoroshev
  normal form, we demonstrate that only a small fraction of the terms
  of the associated optimal remainder provide meaningful contributions
  to the evolution of the action variables.  (ii) We provide rigorous
  analytical approximations to the Melnikov integrals of terms with
  stationary or quasi-stationary phase.  Applying our model to an
  example of three degrees of freedom steep Hamiltonian provides the
  speed of Arnold diffusion, as well as a precise representation of
  the evolution of the action variables, in very good agreement (over
  several orders of magnitude) with the numerically computed one.
\end{abstract}

\section{Introduction}

Fundamental problems of Physics are often modeled with small
Hamiltonian perturbations of integrable systems. For example, the
problem of the stability and long--term evolution of the
Solar System can be modeled in terms of perturbations to 
Kepler's motion of each planet under the gravity of the Sun.  Similar
perturbative approaches are employed in some of the most classical
problems of Mechanics appearing from the microscopic scale (e.g. the
semi-classical treatment of atomic and molecular dynamics) up to the
astronomical one (e.g. solar systems and galaxies). The above and other
important applications, as e.g. in plasma and accelerator physics, or
statistical mechanics, have rendered Hamiltonian
near-integrable systems a fundamental topic in
physics (see \cite{MacKayMeiss87} for a collection of basic
papers and reviews in this field).

One of the most interesting questions in near-integrable systems is
the long--term fate of trajectories which belong to the so-called
`Arnold web'. Following the pioneering work of V.I. Arnold
\cite{Arnold64}, the Arnold web is understood as an intricate in shape
and connected set in phase--space which contains chaotic
trajectories. The Arnold web is tightly related to the existence, in
phase--space, of a corresponding `web of resonances', i.e., domains
where the trajectories undergo near-oscillatory motions with a
commensurable set of frequencies. `Arnold diffusion' is a
theoretically predicted phenomenon, according to which a trajectory
with initial conditions within the distorted separatrices of the
resonances undergoes slow chaotic diffusion. When the number $n$ of
the degrees of freedom is equal to 3 or larger, such diffusion renders
possible, in principle, to connect every part of the Arnold web within
sufficiently long times.  Let us therefore consider a $n$-degree of
freedom Hamiltonian of the form:
\begin{equation}
H_\epsilon(I,\varphi)=H_0(I)+\epsilon f(I,\varphi)
\label{hamilt}
\end{equation}
where $(I,\varphi)\in A\times {\Bbb T}^n$ are action-angle variables,  
$A\subseteq {\Bbb R}^n$ is open bounded, the integrable approximation 
$H_0$ and the perturbation $f$ are real analytic, 
$\epsilon$ is a small parameter. The problem we address in this
paper is the following: 
\vskip 0,4 cm
\noindent
    {\it {\it PROBLEM 1:} For given $H_0,f$, small $\epsilon> 0$, and
      $I_*\in A$ such that $\ell\cdot \nabla H_0(I_*)=0$ for a
        unique $\ell \in {\Bbb Z}^n\backslash 0$ (with its multiples),
      provide a formula which gives the maximum speed of the drift
      along the resonance $\ell \cdot \nabla H_0(I)=0$ (averaged on
      time intervals $T$ longer than $1/\epsilon$) among all the
      solutions of Hamilton's equations with initial conditions
      $I(0),\varphi(0)$ with $\varphi(0)\in {\Bbb T}^n$ and $I(0)$ in
      the ball $B(I_*,C\sqrt{\epsilon})\subseteq A$ of center $I_*$
      and radius $C\sqrt{\epsilon}$, with some $C>0$.}
\vskip 0,4 cm
\noindent
{\bf Remarks:} 

\begin{itemize}

\item[(i)] For special choices of $H_0$, $f$, and 
of the resonant vector $\ell\in {\Bbb Z}^n\backslash 0$, the previous problem 
has a simple solution. In fact, Nekhoroshev provided a class of 
quasi-integrable Hamiltonian systems with variations of the actions of 
order 1 already on times of order $1/\epsilon$ which can be
explicitly computed with a simple quadrature.

\item[(ii)] For $n=2$ and $H_0$ iso--energetically non--degenerate, the KAM
theorem provides a topological obstruction to the drift along 
the resonances of the system, so the previous problem is not interesting.

\item[(iii)] For $H_0$ satisfying a transversality condition, called
  by Nekhoroshev ``steepness'', the stability time of the action
  variables improves dramatically to an exponential order in
  $1/\epsilon$ \cite{N1,N2}: precisely, there exist positive constants
  $a$, $b$ and $\epsilon_0$ such that for any $0\leq \epsilon
  <\epsilon_0$ the solutions $(I(t),\varphi(t))$ of the Hamilton
  equations of $H_\epsilon(I,\varphi)$ satisfy
 \begin{equation}
|I(t)-I(0)|\leq \epsilon^{b}\ \ {\rm for}\ \ 
|t|\ \leq \  T_N:=\frac{1}{\epsilon}\ 
\exp\Big({\frac{1}{\epsilon^{a}}}\Big)  .
\label{statab}
\end{equation}
According to Nekhoroshev's theorem, any large drift of the action variables
needs time intervals longer than the exponentially long--time $T_N$. 
For systems with $n\geq 3$ satisfying the 
hypotheses of Nekhoroshev's theorem, proving the existence of orbits with 
variations of the actions of order 1 in some  suitable long time  
for any small value of $\norm{\epsilon}$, is highly
non trivial, and these are the conditions under which  
Problem 1 is interesting and, up to now, unsolved.

\item[(iv)] We do not provide here a rigorous solution to Problem 1,
  but we provide formulas which match the very slow drifts observed in
  numerical experiments. These formulas are obtained by combining: --
  a semi-analytic argument including the computer assisted computation
  of normal forms, whose coefficients are provided in floating point
  arithmetics, -- a rigorous approximation of the Melnikov integrals
  using methods of asymptotic analysis based on the so called
  stationary-phase approximation, -- a random--phase assumption used
  in the Melnikov approximation. We call our approach
  'semi-analytical', since it combines rigorous results (in the
  stationary-phase method, see Section 3) with ones based on the
  numerical (computer-assisted) computations of the Nekhoroshev normal
  forms. Whether these ideas can be transformed into a fully rigorous
  argument is a question beyond the purpose of this paper.

\item[(v)] The estimated speed of drift along a resonance expected
  from the solution of Problem 1 should depend on $\epsilon$, $I_*$
  and $\ell$.  Of course, close to $I_*$ the resonance $\ell\cdot
  \nabla H_0(I)=0$ may intersect an infinite number of other
  resonances $\tilde \ell\cdot \nabla H_0(I)=0$ with $\tilde\ell\in
  {\Bbb Z}^n\backslash 0$ independent on $\ell$, and solutions with
  initial conditions close to $I_*$ may leave the resonance $\ell\cdot
  \nabla H_0(I)=0$ and drift along different resonances, possibly of
  different multiplicities.  Problem 1 concerns only the orbits which
  drift along the fixed resonance $\ell\cdot \nabla H_0(I)=0$.

\item[(vi)] For systems of $n=3$ degrees of freedom a solution of
  Problem 1 gives the opportunity to compare the time needed to
  diffuse along the resonances of multiplicity 1 with the stability
  time $T_N$ of the Nekhoroshev theorem. In fact for $n=3$
  distant points of the action--space on the same energy level are
  connected through paths of the Arnold web which are mostly contained
  in resonances of multiplicity one, where Problem 1 is
  applicable. The resonances of multiplicity two are just at the
  points of intersection of the resonances of multiplicity one. The
  transit of the orbits through resonances of multiplicity two,
  the so--called 'large gap problem', is one of the hardest
  theoretical difficulties in rigorously proving the existence of
  Arnold diffusion. Numerical studies, instead, provide
    overwhelming evidence for the existence of such transits, see
  \cite{GLF05,GLF09}, while the key question regarding the
    quantification of Arnold diffusion is Problem 1.

\item[(vii)] Throughout this paper, and  mostly in Section
  \ref{numsection}, we compare our semi-analytic solution of Problem 1
  with numerical experiments. The long-term behaviour of Hamiltonian
  systems, including Arnold diffusion, can be numerically investigated
  with symplectic integrators (see \cite{BG94,Nei84,GV18}). In fact,
  depending on the order of the integration scheme, every step
  $\phi^\tau$ of the integrator is exponentially close, with
  exponential factor $-1/\tau$, to the exact Hamiltonian flow of a
  modified Hamiltonian
  $$
  K_\epsilon(I,\varphi) = H_\epsilon(I,\varphi)+\tau^\nu W
  (I,\varphi;\epsilon)~,
  $$
  where the integer $\nu$ and $W$ both depend on the integration
  scheme.  Therefore, for suitably small $\tau$ the spurious term $
  \tau^\nu W (I,\varphi;\epsilon)$ is just a perturbation of the
  original Hamiltonian, and the exponential factor becomes negligible
  with respect to any observed diffusion.

\item[(viii)] While the KAM and Nekhoroshev theorems, as well as the
  examples of Arnold diffusion (and also Problem 1), are usually
  formulated for quasi-integrable hamiltonians (\ref{hamilt}), many
  quasi-integrable systems of interest for Physics and Celestial
  Mechanics are characterized by degeneracies and singularities of the
  action variables which introduce additional complications. Many
  researches recovered the proofs of the KAM and Nekhoroshev
  theorems, as well as of the existence of hyperbolic tori, also for the
  cases of interest for Celestial Mechanics, see for example,
  \cite{ChGall,Benfagu,gioretal89,GLS, CP,P16,Delsh18}.

\end{itemize}
\vskip 0,2 cm Problem 1 is an applicative spin-off of the problem of
Arnold diffusion, which started with the fundamental paper published
by Arnold in 1964 \cite{Arnold64}, first providing a quasi--integrable
Hamiltonian system with non trivial long--term instability.  Since
Arnold's pioneering paper, a rich literature has appeared on attempts
to prove of existence of Arnold diffusion for more general
quasi--integrable Hamiltonian systems, called, in the context of
Arnold diffusion, {\it a priori stable} systems. A simpler, albeit
still highly non trivial, case is the one of {\it a priori unstable}
systems. In the latter case, the existence of diffusing motions has
been proved using different models and techniques, including Mather's
variational methods, geometrical methods and the so-called separatrix
and scattering maps (among the rich literature see
\cite{ChGall,BeBiBo,Delshetal06,trechev04,ChengYan,BKZ,GideaMarco,KZ,KZ15,Delshetal16,CapGi}
and references therein).

Due to the long timescales involved, also numerical or experimental
observations of Arnold diffusion are hard to achieve. Already few
years after the first numerical detection of chaotic motions
\cite{henheil}, the long-term instability in Hamiltonian systems was
discussed from both an analytical and numerical point of view in
\cite{Chir}. However, only modern computers rendered possible to
simulate the phenomenon in simple physical models. In the last
decades, diffusion through the resonances has been clearly detected
\cite{Laskar93,EVC98,kaneko,gc04,LGF03,GLF05,FGL05,GLF11,gc04,efthy,EfthyHarsoula,TGLF,GL13,GL16}. Then,
in the series of papers \cite{LGF03,GLF05,FGL05,GLF11,GL16}, diffusion
of orbits has been also detected for values of the perturbation
parameters so small that the set of resonant motions has the structure
of the Arnold web embedded in a large volume of invariant tori (the
distributions of resonances and tori being computed numerically with
chaos indicators \cite{FGL00,LGF03,GL13}). In these experiments, the
instability was characterized by diffusion coefficients decreasing
faster than power laws in $\epsilon$, compatibly with the exponential
stability result of Nekhoroshev's theorem. This was confirmed by a
direct comparison of the numerical diffusion coefficient with the size
of the optimal remainder of the Nekhoroshev normal form
in~\cite{efthy}.

In this paper we propose a semi-analytic solution to Problem 1 which is 
obtained through the following steps: 
\vskip 0,4 cm
\noindent
(a) Given $\epsilon$ and $I_*$ we construct
a computer assisted normal form adapted to the local resonance
properties at $I_*$ up to an optimal normalization order, by following
the construction of normal forms which appears within the proof of 
the Nekhoroshev theorem. The computer assisted construction of normal 
forms is mandatory, 
since our purpose is to compare the predicted values of 
the drifts with the numerically observed ones. As it is well known
 (see \cite{CellChier95,CellChier97} for the KAM
theorem, and \cite{CF96,GLS} for the Nekhoroshev theorem) 
purely analytic estimates which do not use computer assisted methods 
are highly unrealistic. 
\vskip 0,2 cm
\noindent
(b) We represent the variation of the actions along the resonance with
Melnikov integrals defined from the normal form constructed as
indicated in (a). Since the remainders of these normal forms are
represented as expansions of millions of very small terms, the
variation of the actions is represented as a sum of millions of
Melnikov integrals, which have to be computed in order to solve
Problem 1.  The problem becomes prohibitive if the goal is to
  maximize the result with respect to some variables in order to
  compute the orbits with largest instability. To overcome this
  difficulty, we require an analytic method that allows to
  descriminate between the terms of the remainder associated with
  large contribution to the variations of the actions and those of
  negligible contribution, therefore reducing the total amount of
  terms to consider.

\vskip 0,2 cm
\noindent
(c) We represent the Melnikov integrals with a method from asymptotic
analysis, the so-called {\it method of the stationary-phase} (see
\cite{BleHan}).  In fact, for quasi-integrable systems, the Melnikov
integrals can be reformulated as integrals with a rapidly oscillating
phase, and the computation of the critical points of this phase
provides an estimate of the integral.  We find that only the Melnikov
integrals whose phase either 1) has critical points, or 2) the
derivative of the phase with respect to the slow angle variable of the
resonance is suitably small, provide major contributions to the Arnold
diffusion. We call the corresponding terms in the remainder
stationary or quasi-stationary, respectively.  The Melnikov integrals
whose phase is neither stationary nor quasi-stationary represent the
large majority of terms, and their cumulative contribution to the
Arnold diffusion is negligible with respect to the cumulative
contribution of the stationary or quasi-stationary terms. Therefore,
we provide a rigorous criterion to select, from the millions of
harmonics of the remainder of the Nekhoroshev normal form, a few
thousand ones. All the relevant integrals of Melnikov theory can be
explicitly represented with an asymptotic formula or directly computed
by quadratures. The asymptotic formula, providing the variation of the
actions during a resonant libration, depends on the initial phases
$\varphi(0)$. For all possible values of these phases the formula
represents closely the spread of the actions which is observed with
numerical integrations.
\vskip 0,2 cm
\noindent
(d) Finally, by maximizing with respect to $\varphi(0)$ the
variations of the actions obtained from the Melnikov integrals we
obtain the orbits with largest variation of the actions at each
homoclinic loop, as well as the rare initial conditions whose orbit,
in a sequence of homoclinic loops, have a systematic variation of the
action variables.  Thus we predict which orbits undergo
  the 'fastest' Arnold diffusion which we can observe.
\vskip 0,4 cm

From (a),~(b),~(c),~(d) we have a qualitative and quantitative
description of the drift along the resonances of multiplicity one.
The qualitative picture of the diffusion is in agreement with the idea
having its roots in Chirikov's fundamental paper \cite{Chir} and
recently recovered e.g. in \cite{CEGM}, namely that the
diffusion along a resonance is not uniform in time, but it is produced
by impulsive `kicks' or `jumps' at every homoclinic loop, see
\cite{Bencarfas97,Delshetal97,Rudwig1998}. The new quantitative
analysis allows us to determine the frequency of occurrence and
amplitude of these jumps as the resonant angle becomes critical for
some Melnikov integrals; also, we are able to select the initial
conditions whose orbits have the fastest Arnold diffusion. Therefore,
for given values of $\epsilon$, we are able to predict the minimum
timescales needed to observe long--term diffusion along any
  single resonance.

The paper is organized as follows. In Section 2 we define the Melnikov
integrals from the normal forms of Nekhoroshev theory.  In Section 3
we provide rigorous asymptotic representations of the Melnikov
integrals using the method of stationary phase approximation. In
Section 4 we present a semi-analytic solution to Problem 1. Section
\ref{numsection} is devoted to a numerical demonstration of the theory
presented in Sections 2,~3,~4.

\section{Nekhoroshev normal forms and Melnikov integrals} 

The long--term dynamics of the quasi--integrable Hamiltonian
(\ref{hamilt}) is traditionally studied using the averaging method. In
the refined version of the method defined within the proof of
Nekhoroshev's theorem, for a $d$--dimensional lattice
$\Lambda\subseteq {\Bbb Z}^n$ defining the resonance
$$
{\cal R}_\Lambda=\{ I\in A:\ \ \ell \cdot \nabla H_0(I)=0 , \ \ \forall \ell\in \Lambda\}  ,
$$
one constructs a canonical transformation
$$
(S,F,\sigma,\phi)={\cal C}(\Gamma^{-T}I,\Gamma \varphi)
$$ 
defined in a suitable resonant domain $D_\Lambda\times {\Bbb T}^n$, where
\begin{itemize} 
\item [- ] $\Gamma$ is a matrix with $\Gamma_{ij}\in {\Bbb Z}$ 
and $\det \Gamma=1$, that defines a linear canonical 
map (see~\cite{bengall86})
\begin{equation}
(\tilde S,\tilde F,\tilde \sigma,\tilde \phi)=
(\Gamma^{-T}I,\Gamma \varphi)
\label{maps}
\end{equation}
and conjugates $H_0(I)$ to $h(\tilde S,\tilde F)$ such that the
resonance ${\cal R}_\Lambda$ is transformed into
$$
\tilde {\cal R}=\{ (\tilde S,\tilde F)\in \Gamma^{-T}A:\ \ {\partial  \over 
\partial \tilde S_j}h(\tilde S,\tilde F)=0 , \ \ \forall j=1,\ldots ,d \}~.
$$
Equivalently, in the new variables the resonant lattice $\Lambda$ is 
transformed into the lattice generated by $e_1,\ldots ,e_d$,
($e_1,\ldots ,e_n$ denotes the canonical basis of ${\Bbb R}^n$).

\item[ -] $\sigma\in {\Bbb T}^d,\phi\in {\Bbb T}^{n-d}$ are angles conjugate
to the actions $S\in {\Bbb R}^d,F\in {\Bbb R}^{n-d}$.

\item[ -] $D_\Lambda$ is a domain whose definition depends
both on the resonant lattice $\Lambda$ and on $\epsilon$.

\item[ -] ${\cal C}$ is a near to the identity transformation and,
  when composed with (\ref{maps}), conjugates $H_\epsilon$ to the
  Nekhoroshev normal form Hamiltonian
\begin{equation}
H_{\epsilon,\Lambda}=h(S,F)+\epsilon f_\Lambda(S,F,\sigma)+
 r_\Lambda(S,F,\sigma,\phi)~,
\label{hamiltlambda}
\end{equation}
where the remainder $r_\Lambda$ has norm bounded by a factor 
exponentially small with 
respect to $-1/\epsilon^a$ (see \cite{N1,P,Lochak92,LochakN92,GCB16}
for precise definitions and statements).  
\end{itemize}
The integer $d\in {1,\ldots ,n-1}$ is called the multiplicity of the resonance.

Although the proof of Nekhoroshev's theorem grants the existence
of normal forms (\ref{hamiltlambda}) for suitably small values $0\leq
\epsilon\leq\epsilon_0$ (see (\ref{statab})), the precise value
  of the threshold parameter $\epsilon_0$ is believed to be largely
underestimated  by the general proofs, while the Fourier coefficients of
$f_\Lambda,r_\Lambda$ are estimated in norm, but not explicitly
provided. Both problems can be overcomed by
constructing the normal forms (\ref{hamiltlambda}) with computer
assisted methods \cite{gioretal89,CF96,gsk,efthy}.  We call
Hamiltonian normalizing algorithm (HNA) a computer-algebraic
implementation which provides the coefficients of the Nekhoroshev
normal form (\ref{hamiltlambda}). We use the HNA introduced in
\cite{efthy}, which normalizes quasi-integrable Hamiltonians
$H_\epsilon$. The HNA is constructed by composing $N$ elementary
transformations; the input of the HNA is the Hamilton function, a
resonance lattice $\Lambda$, a domain $D\times {\Bbb T}^n$ where the
transformation if defined; the output of the HNA is a canonical
transformation $(F,S,\sigma,\phi)={\cal C}^N(I,\varphi)$ and a normal
form Hamiltonian
\begin{equation}
H^N= h(F,S)+\epsilon f^N (F,S,\sigma)+ r^N (F,S,\sigma,\phi)\label{HN}~,
\end{equation}
conjugate to $H_\epsilon$ by ${\cal C}^N$. 
The remainder $r^N$ is provided as a Taylor-Fourier series
\begin{equation}
r^N = \sum_{m\in {\Bbb N}^d}\sum_{\nu \in {\Bbb Z}^d} \sum_{k\in {\Bbb
    Z}^{n-d}}r^{m}_{\nu,k}(F-F_{*})(S_1-S_{*,1})^{m_1} \cdots
(S_d-S_{*,d})^{m_d} e^{i\nu \cdot \sigma+ik\cdot \phi}
\label{remainder}
\end{equation}
expanded at a suitable $(F_{*},S_*)$, with computer-evaluated truncations
involving a large number (typically $\sim 10^7,10^8$) of terms.  $N$
is chosen so that $\Norm{r^1} > \ldots > \Norm{r^N}$ and
$\Norm{r^{N+1}}> \Norm{r^{N}}$, thus the normal form is called
optimal, $r^N$ the optimal remainder and $N$ the optimal normalization
order (the norm definitions in the selected domain are as in
\cite{efthy}).

If we artificially suppress the remainder $r_\Lambda$ in
Eq.~(\ref{hamiltlambda}), or $r^N$ in (\ref{HN}), we obtain an
exponentially small perturbation of the original Hamiltonian which
possibly exhibits chaotic motions due to homoclinic and heteroclinic
phenomena (for $d>1$), but in which the actions $F_j$, which we
call 'adiabatic', remain constant in time. Therefore, in the flow
of the complete Hamiltonian, any long--term evolution of the adiabatic
actions is due to the accumulation of the effects of the very small
remainder on very large times. In particular, the adiabatic actions
$F_j$ have a long-term variation, representing the drift along
the resonance, bounded for an exponentially long time by
\begin{equation}
\norm{F_j(t)-F_j(0)}\leq \norm{t}
\Norm{{\partial r_\Lambda\over \partial 
\varphi_j}}~.
\label{apriori}
\end{equation}
The {\it a priori} estimate (\ref{apriori}) obtained from the Nekhoroshev 
normal form (\ref{hamiltlambda}) provides an upper bound to the average 
variation of the adiabatic actions; establishing {\it lower bounds} to 
$\norm{F_j(t)-F_j(0)}$ is the fundamental brick in the theory of Arnold
diffusion.
\vskip 0,4 cm

From now on we focus our discussion on resonances of 
multiplicity $d=1$. Denoting by $(S,\sigma)$ the resonant action-angle pair,  
we first consider the dynamics of the approximated normal form 
which is obtained from (\ref{HN}) just by dropping the remainder 
$r^N (F,S,\sigma,\phi)$:
\begin{equation}\label{HNappr} 
\overline H^N= h(F,S)+\epsilon f^N(F,S,\sigma) .
\end{equation} 
Since the corresponding Hamiltonian $\overline H^N$ depends only on
one angle, it is integrable, and we represent its motions as
follows. Following \cite{bengall86}, we first expand $\overline H^N$
at $(F_*,S_*)$ identifying the center of the resonance, precisely such
that
\begin{equation}\label{centerFS}
{\partial h\over \partial S}(S_*,F_*)=0 .
\end{equation}
We obtain
\begin{equation}\label{HB0}
\overline H^N= \overline H_0+...\ \ ,\ \ \overline H_0=\omega_*\cdot \hat F
+ {A\over 2}\hat S^2 + \hat S  B\cdot \hat F +
{1\over 2} C\hat F\cdot \hat F + \epsilon v(\sigma)~,
\end{equation}
where $\hat F=F-F_*$, $\hat S=S-S_*$, and the Hamiltonian $\overline
H_0$ is represented using a number $A\in {\Bbb R}$, two vectors
$\omega_*,B\in {\Bbb R}^{n-1}$, a square matrix $C$ and a function
$v(\sigma)$, all these quantities depending parametrically on
$S_*,F_*$.

The actions $\hat F$ are constants of motion for the Hamiltonian flow of
$\overline H^N$ as well as of $\overline H_0$. 
We parameterize by
\begin{equation}
\hat S= \sqrt{\epsilon}s_\alpha(\sigma) :=  
\pm \sqrt{\epsilon}\sqrt{{2\over \norm{A}}(M(1+\alpha)-v(\sigma))}~,
\label{hats}
\end{equation}
the level curves of $\overline H_0$ for $\hat F=0$, where 
$$
M= \max_{\sigma\in [0,2\pi]} v(\sigma)  ,
$$
and $\alpha$ is a convenient label for the energy levels of 
$\overline H_0$. For any $\alpha\ne 0$, we denote by $T_\alpha$ the period of 
the corresponding solutions of Hamilton's equations under $\overline H_0$ (for 
$\hat F=0$). We also define
$$
\bar M= \min_{\sigma\in [0,2\pi]} v(\sigma)~,
$$
and, without loss of generality, we assume $\bar M\leq 0$, $M>0$. 

When considering the solutions $(F(t),S(t),\sigma(t),\phi(t))$ of
Hamilton's equations under the complete Hamiltonian (\ref{HN}), the
adiabatic actions $F_j$ can have a slow evolution forced by the
remainder $r^N$, whose variation $\Delta F_j(T)=F_j(t)-F_j(0)$ in a
time interval $[0,T]$ is given by
\begin{equation}
\Delta F_j(T) =-\sum_{m,\nu,k} \int_0^{T}  i k_j r^m_{\nu,k}(F(t))\hat S(t)^m 
e^{i\nu \sigma(t) + ik\cdot \phi(t)}dt
:= \sum_{m,\nu,k} \Delta F_{j,T}^{m,\nu,k}~~.\label{sumF0full}
\end{equation}
According to the 
well known Melnikov approach (see \cite{Chir} for 
a review) we approximate $\Delta F_j(T)$ with
\begin{equation}
-\sum_{m,\nu,k} \int_0^{T}  
i k_j r^m_{\nu,k}(F_*){\hat S^0(t)}^m 
e^{i\nu \sigma^0(t) + ik\cdot \phi^0(t)}dt~~,
\label{sumF0}
\end{equation}
obtained by replacing the solution $(F(t),S(t),\sigma(t),\phi(t))$ in the 
integrals with
$$
(F_*,S^0(t),\sigma^0(t),\phi^0(t))=(F_*,S_*,0,0)+
(0, \hat S^0(t),\sigma^0(t),\phi^0(t))
$$
where $(0,\hat S^0(t),\sigma^0(t),\phi^0(t))$ is a fixed solution
of Hamilton's equations under $\overline H_0$ (see remark (ix)
  below). Finally, by changing the integration variable from $t$ to
$\sigma=\sigma^0(t)$, we have
\begin{equation}
\Delta F_{j,T}^{m,\nu,k}\simeq 
\Delta^0 F_{j,T}^{m,\nu,k}  := 
- i k_j {{r^m_{\nu,k}(F_*) \epsilon^{m-1\over 2}\over A}e^{ik\cdot\phi(0)}}
\int_0^{\sigma^0(T)}\hskip -0.6cm  [s_\alpha(\sigma)]^{m-1} 
e^{i \theta(\sigma)}d\sigma
\label{melnikint}
\end{equation}
where the phase $\theta(\sigma)$ is defined by:
$$
\theta(\sigma)= {{\cal N}\sigma +{\Omega\over A\sqrt{\epsilon}}
\int_0^{\sigma} {dx\over s_\alpha(x)} } 
$$
with 
\begin{equation}
{\cal N}= \nu +k\cdot B/A,~~~~ \Omega =k\cdot \omega_*~~~. 
\label{calnomega} 
\end{equation}
{\bf Remark} (ix). Usually, Melnikov approximations are introduced to
compute the splittings of stable-unstable manifolds, so that integrals
like (\ref{sumF0full}) are approximated by choosing
$(\hat{S}^{0}(t),\sigma^{0}(t),\phi^{0}(t))$ to be the solution of the
approximate normal form $\overline H_0$ corresponding to a separatrix
homoclinic loop ($\alpha=0$ in our notation). Our method exposed below
differs from the usual Melnikov approach since, in order to find the
orbits which diffuse in shorter time along the resonance, we evaluate
the integrals for a solution
$(\hat{S}^{0}(t),\sigma^{0}(t),\phi^{0}(t))$ which is suitably close
to, but {\it not exactly on} the separatrix, precisely
$\alpha\sim\Norm{r^N}$, with finite period $T_\alpha$.

\section{Stationary phase approximation of Melnikov integrals}

\subsection{\label{statia} The principle of stationary phase}
While formulas (\ref{sumF0full}) or (\ref{sumF0}) allow, in principle, 
to compute the time evolution of the adiabatic actions during consecutive  
homoclinic loops along the resonance, the evaluation of the sums over 
millions of remainder terms $r^m_{\nu,k}$ is hardly tractable in practice. 
We find that most of these terms (including some of 
the largest in norm) contribute very little to the sum 
(\ref{sumF0}). This fact can be explained by invoking methods of 
asymptotic analysis inspired by the so--called principle of the stationary 
phase (PSP hereafter). In its classical formulation
 (e.g., see \cite{BleHan}) the principle concerns
the asymptotic behaviour of the 
parametric integrals
\begin{equation}
I_\lambda = \int_a^b \eta(\sigma ) e^{i \lambda  \Phi(\sigma )} d\sigma ~,
\label{IPSP}
\end{equation}
when the parameter $\lambda$ is large.  
With mild conditions on the amplitude function $\eta(\sigma)$, 
we have the following cases:
\begin{itemize}
\item[(A)]The phase $\Phi(\sigma )$ has no stationary points, i.e. 
$\Phi(\sigma )\ne 0$ for all $\sigma\in [a,b]$, for large $\lambda$ we have
\begin{equation}
I_\lambda \sim {\eta(\sigma )\over \lambda \Phi'(\sigma )}
e^{i(\lambda \Phi(\sigma ))}{\Bigg \vert}_a^b ,
\label{est1}
\end{equation}
where the neglected contributions are of order smaller than $1/\lambda$. 

\item[(B)] The phase $\Phi(\sigma )$ has a non--degenerate 
stationary point $\sigma_c\in (a,b)$, i.e. $\Phi'(\sigma_c)=0$ and 
$\Phi''(\sigma_c)\ne 0$.
Then, if $\eta(\sigma_c)\ne 0$, for large $\lambda$ we have
\begin{equation}
I_\lambda \sim \eta(\sigma_c) e^{i\left(\lambda \Phi(\sigma_c)\pm {\pi\over 4}\right )} 
\sqrt{2\pi \over \lambda \norm{\Phi''(\sigma_c)}}~,
\label{est2}
\end{equation}
where the $\pm$ is chosen according to the sign of $\Phi''(\sigma_c)$. 
If there are more stationary points in $(a,b)$, we must sum 
all the corresponding terms. 

\item[(C)]  The phase $\Phi(\sigma )$ has a degenerate
stationary point $\sigma_c\in (a,b)$, i.e. $\Phi'(\sigma_c)=0,\Phi''(\sigma_c)= 0$ and we assume $\Phi'''(\sigma_c)\ne 0$. Then, if 
$\eta(\sigma_c)\ne 0$, for large $\lambda$ we have
\begin{equation}
I_\lambda \sim \eta(\sigma_c) e^{i \lambda \Phi(\sigma_c)} \sqrt{3}\Gamma(4/3)
\left ({6 \over \lambda \norm{\Phi'''(\sigma_c)}}\right )^{1\over 3}  . 
\label{est3}
\end{equation}
\end{itemize}

\subsection{Heuristic discussion of the PSP for Melnikov integrals} 

Let us consider the Melnikov integrals (\ref{melnikint})
\begin{equation} \label{eq:calI}
{\cal I}=
\int_0^{\sigma^0(T)}\hskip -0.6cm  [s_\alpha(\sigma)]^{m-1} e^{i \theta(\sigma)}d\sigma
\end{equation}
and identify the oscillating phase with 
$\theta(\sigma):= \lambda \Phi(\sigma)$. 
Since the derivative of $\theta(\sigma)$
$$
 {d\theta\over d\sigma}= {\cal N} +{\Omega\over A\sqrt{\epsilon}}
{1\over s_\alpha(\sigma )}
$$
is, in principle, divergent for $\epsilon$ going to zero, we will
justify in the next sub-section the use of the method of the
stationary phase by showing how to identify the large parameter
  $\lambda$ in terms of specific parameters entering into the calculus
  of~\eqref{eq:calI}.  According to this idea, the Melnikov integrals
whose oscillating phase $\theta(\sigma)$ has critical points are
expected to be dominant over those whose oscillating phase has
no critical points. However, we find that the asymptotic
behaviour of the Melnikov integrals is more complicated than the
behaviour of the integrals (\ref{IPSP}) thus rendering necessary
  to use a refinement of the stationary phase method: Specifically,
we need to consider also a case which is intermediate between (A) and
(B), called hereafter the quasi-stationary case, produced by the
disappearance of couples of non-degenerate critical points
$\sigma^1_c,\sigma^2_c$ after they merge into a degenerate critical
point. The quasi-stationary case represents a transition between the
stationary and the non-stationary case, which is not considered in the
usual formulations of the PSP method.  To be more precise,
depending on the values of $m,\nu,k$, we will consider three cases:
\begin{itemize}
\item[(I)] {\it the phase $\theta(\sigma)$ is 'fast' for 
all $\sigma\in [0,2\pi]$, i.e. we have $\norm{\theta'(\sigma)} > \gamma$,
with $\gamma$ a large parameter to be defined later.} 

In this case the integral in (\ref{melnikint}) is estimated 
smaller than order $\lambda^{-1}$, see Eq.~(\ref{est1}), and 
we will assume that the  contribution  of $\Delta F_{j,T}^{m,\nu,k}(T)$ 
to the series expansion in Eq.~\eqref{sumF0full} can be neglected.  

\item[(II)] {\it the phase $\theta(\sigma)$ is 'slow' only close to 
non-degenerate critical points $\sigma_c$, provided they are 
distant enough with respect to $1/\lambda^{1\over 3}$.}

In this case, by invoking the PSP 
(see (\ref{est2})), the integrals
  $\int_0^{\sigma}[s_\alpha(\sigma)]^{m-1} e^{i \theta(\sigma)}d\sigma$ 
  are approximated by the sum on all the
  stationary points $\sigma_c\in [0,\sigma]$ of contributions ${\cal
    I}(\sigma_c)$ defined by
\begin{equation}
{\cal I}(\sigma_c)= [s_\alpha(\sigma_c)]^{m-1} e^{i \theta(\sigma_c)\pm i
  {\pi\over 4}} {\sqrt{2\pi} \over \sqrt{ {\norm{\Omega} \over
      A^2\sqrt{\epsilon}} {\norm{v'(\sigma_c)} \over
      \norm{s_\alpha(\sigma_c)}^3} }}\label{ic1}~,
\end{equation}
where the $\pm$ depends on the sign of $v'(\sigma_c)$. The above
formula is valid if $\theta''(\sigma_c) \ne 0$. Consequently, we obtain 
(see Lemma 1):
\begin{equation}
\Delta^0 F_{j,T}^{m,\nu,k} \sim  
- i k_j {r^m_{\nu,k}(F_*) \epsilon^{m-1\over 2}\over A}
e^{ik\cdot \phi(0)}\hskip -0.4 cm\sum_{\sigma_c\in [0,\sigma^0(T)]}
\hskip -0.4 cm{\cal I}(\sigma_c) .
\label{melnikint2}
\end{equation}

\item[(III)] {\it the phase $\theta(\sigma)$  is quasi-stationary, 
i.e. $\norm{\theta'(\sigma)}\leq\gamma$ in an interval of 
size of order $1/\lambda^{1\over 3}$ (notice that this condition
can occur in absence of critical points, or in presence of two very 
close non-degenerate critical points or of 
one degenerate critical point).}
  \vskip 0.2 cm
  While in this case we cannot directly apply (A), (B),
(C), we will obtain an asymptotic formula for the integrals stemming
from formula (C) (see Lemma 2).
\end{itemize}

{\bf Remarks:} 
\begin{itemize}

\item[ (x)] from estimates (A) and (B), any individual integral estimated
using (\ref{est2}) is of order $1/\sqrt{\lambda}$, larger with respect to the 
integrals estimated using (\ref{est1}) which are of order $1/\lambda$. 
Moreover, in the non--stationary case (I) the integrals in (\ref{est1})
are estimated only by the difference of a function computed at the 
border values $a,b$. Since Arnold diffusion is produced by the variations of 
$F_j$ through a sequence of circulations or librations, the border 
values of the sequence may cancel (as a matter of fact they may cancel 
only partially, 
since from a libration/circulation to the next one there can be small
variations of $\alpha$ and a change of the fast phases $\phi(0)$
in the factor multiplying the integral in (\ref{melnikint})).


\item[(xi)] 
The practical classification of the integrals in 
one of the categories (I), (II), or (III) will be done by a fast algorithmic 
criterion (see below), based only on each term's integer labels $m,\nu,k$. 
Since for the large majority of $\nu,k$ the phase satisfies (I) 
(see Table~1 of the numerical examples), 
we have a criterion to select the few harmonics ($\sim$few in 1000) 
belonging to (II) and (III), hence, producing the dominant terms in the 
time evolution of $F_j$.
 
\item[(xii)]
  Despite being more complicated than (II), the inclusion
  in the computation of the quasi--stationary terms (III) is
  essential, since these individual contributions can be as large as
  those of (II) and quite often we find algebraic near-cancellations
  between terms of the groups (II) and (III) leaving residuals of
  order only few percent of the absolute values of the corresponding
  Melnikov integrals.
 \end{itemize}

\subsection{Rigorous discussion of the PSP for Melnikov integrals} 

In this subsection we state the results which allow us to assign
all the terms in Eq.~\eqref{melnikint}, labeled by the integers
$m,\nu\in {\Bbb Z}$ and $k\in {\Bbb Z}^{n-1}$, into the categories
(I), (II) or (III). We then provide rigorous asymptotic
representations for the integrals of the terms in (II) and
  (III).  To simplify the discussion, we assume $\alpha> 0$; the
modifications needed to represent the case $\alpha<0$ are
straightforward.

We first analyze the conditions on $m,\nu,k$ which imply that 
the corresponding term has a phase with stationary points. To
simplify the analysis we assume mild conditions on the potential 
$v(\sigma)$ in the normal form Hamiltonian 
(\ref{HB0}). The first hypothesis is that $v(\sigma)$ has only one
point of maximum. Up to a translation of 
the angle $\sigma$, we assume it at $\sigma=0$, so that with 
the notations introduced in Section 2, we have $v(0)=M$, implying
\begin{equation}
\theta(\sigma)=  {\cal N}\sigma +{\cal W}{ \sqrt{1-{\bar M\over M}}} 
\int_0^\sigma {dx\over  \sqrt{1+\alpha - {v(\sigma ) \over M}}} ,
\ \ \ \ {\cal W}= \pm {\Omega \over \sqrt{2\norm{A}\epsilon (M-\bar M)}} ~,
\end{equation}
with the sign $\pm$ chosen according to the 
signs of $s_\alpha(\sigma)$ and $A$. We now have the following:
\vskip 0.2 cm
\noindent
{\bf Lemma 1.} { \it Consider a phase $\theta(\sigma )$ defined by the
  labels $m,\nu,k$. Then:
\begin{itemize}
\item[-] If $ {\cal N}\cdot{\cal W}>0$ the phase $\theta(\sigma )$ 
has no stationary points; 
\item[-] If $ {\cal N}\cdot{\cal W}<0$, the 
phase $\theta(\sigma )$ has stationary points if and only if
\begin{equation}
{\sqrt{1-{\bar M\over M}} \over \sqrt{1+\alpha -{\bar M\over M}}} \leq 
{|{\cal N}|\over |{\cal W}|}\leq 
{\sqrt{1-{\bar M\over M}} \over \sqrt{\alpha}}  .
\label{condNWalpha}
\end{equation}
\end{itemize}

Furthermore, suppose that $v(\sigma)$ has only one non-degenerate local maximum 
at $\sigma=0$, one non-degenerate local minimum at $\sigma=\bar \sigma$, 
and $v'(\sigma)\ne 0$ elsewhere. For any given $\Delta_{Max}>\sqrt{1-{\bar M\over M}}$, consider $\epsilon$ suitably small and 
\begin{equation}
\alpha \in \left (0,\min \left (\Norm{r^N},{1-{\bar M\over M}\over 
2^8\Delta^2_{Max}}\right )\right )  .
\label{delmax}
\end{equation}
Then
\begin{itemize}
\item[(i)] If the inequality (\ref{condNWalpha}) is strictly
  satisfied, the phase $\theta(\sigma )$ has two non--degenerate
  critical points $\sigma^1_c,\sigma^2_c$ and for all ${\cal N}=
  -{\cal W}\Delta$ with $\Delta \in \left ({\sqrt{1-{\bar M\over M}}
    \over \sqrt{1+\alpha -{\bar M\over M}}},\Delta_{Max}\right )$, we
  have
\begin{equation}
\int_{0}^{2\pi}[s_\alpha(\sigma)]^{m-1}e^{i\theta(\sigma)}d\sigma=
{\cal I}(\sigma_c^1)a_1({\cal W},{\cal N})+
{\cal I}(\sigma_c^2)a_2({\cal W},{\cal N})~,
\label{rigorous1}
\end{equation}
where ${\cal I}(\sigma)$ has been defined in (\ref{ic1}), and the
  functions $a_1$, $a_2$ satisfy
\begin{equation}
\lim_{\norm{\cal W}\rightarrow +\infty}a_1({\cal W},-{\cal W}\Delta)=1~,
\lim_{\norm{\cal W}\rightarrow +\infty}a_2({\cal W},-{\cal W}\Delta )=1 ~.
\label{limits1} 
\end{equation}

\item[(ii)] If ${|{\cal N}|\over {|{\cal W}|}}$ is equal to the
  lowermost bound of Eq.\eqref{condNWalpha}, the two non-degenerate
  critical points merge into one degenerate critical point
  $\sigma_c=\bar \sigma$.

\end{itemize}

}
\vskip 0.2 cm
\noindent
{\bf Proof of Lemma 1.} Since we have
$$
\theta'(\sigma)= {\cal N}+ {\cal W} \sqrt{1-{\bar M\over M}}{1 \over 
\sqrt{1+\alpha - {v(\sigma ) \over M}}}  ,
$$
if ${\cal N}\cdot {\cal W}>0$ there are no stationary points. 
If ${\cal N}\cdot {\cal W}<0$ the stationary points are the solutions of 
the equation
$$
{\norm{{\cal N}} \over \norm{{\cal W}}}=
{\sqrt{1-{\bar M\over M}} \over 
\sqrt{1+\alpha - {v(\sigma ) \over M}}} ~,
$$
which exist if and only if ${|{\cal N}|/ {|{\cal W}|}}$ satisfies
(\ref{condNWalpha}). By assumption, $v(\sigma)$ has only one
local maximum $\sigma=0$ and one local minimum $\bar \sigma$.  If
(\ref{condNWalpha}) is strictly satisfied, the function
$${\tilde \theta}' =\norm{\cal N}-\norm{{\cal W}} \sqrt{1-{\bar M\over
    M}}{1 \over \sqrt{1+\alpha - {v(\sigma ) \over M}}} ,
$$ has a strict maximum at $\sigma=\bar\sigma$ with ${\tilde
  \theta}(\bar \sigma)>0$ and converges to $-1\over \sqrt{\alpha}$ for
$\sigma$ tending to $0$ or $2\pi$. Therefore there are two values
$\sigma^1_c,\sigma^2_c\in (0,2\pi)\backslash \{\bar\sigma\}$ such that
${\tilde \theta}'(\sigma_c^i)=0$. Then, from
$$
\theta''(\sigma)={{\cal W}\over 2M} \sqrt{1-{\bar M\over M}}
{v'(\sigma ) \over \left (1+\alpha - {v(\sigma ) \over M}\right )^{3\over 2}} ~,
$$
we have $\theta''(\sigma^i_c)\ne 0$, and consequently the two critical points 
are non-degenerate. Instead, when the inequality 
(\ref{condNWalpha}) is satisfied at its lower extremum, 
we have only one critical point $\sigma_c=\bar\sigma$, which is degenerate. 

It remains to prove Eq.~(\ref{rigorous1}). With no loss of generality,
  we consider the case ${\cal N}>0,{\cal W}<0$.  Setting ${\cal N}$
within the phase $\theta(\sigma)$ with ${\cal N}= -{\cal W}\Delta=
\Delta \norm{\cal W}$, we obtain $\theta(\sigma)= \norm{\cal
  W}\Phi(\sigma)$ with
$$
\Phi(\sigma)=\left (\Delta\sigma -
\sqrt{1-{\bar M\over M}}\int_0^{\sigma} {dx\over \sqrt{1+\alpha-
{v(\sigma)\over M}}} \right )  .
$$
Since, depending on the values of $m$, the 
integral
$$
\int_{0}^{2\pi}[s_0(\sigma)]^{m-1}e^{i\theta(\sigma)}d\sigma
$$
may not be smooth at $\sigma=0,2\pi$, we use the technique called
{\it neutralization} of the extremals. Precisely, choose a small 
$\mu>0$, depending possibly on the given $\Delta_{Max}$, but 
independent of $\norm{\cal W}$ and $\Delta$. 
In the following we denote by $k_1,k_2,\ldots$~suitable 
constants which do not depend on $m,{\cal W},\Delta,\epsilon$, 
while they may depend on $\mu,\Delta_{Max}$.

We first prove that in the hypothesis of the Lemma there exists 
a small $\mu$ such that both  critical points $\sigma^j_c$ are in 
$(\mu,2\pi-\mu)$, and for any $\sigma\in [0,\mu]$, we have
\begin{equation}
\norm{\Phi'(\sigma)} \geq k_1\ \   , \norm{\Phi'(\sigma)s_\alpha(\sigma)}\geq k_1
 ~.\label{phiprimo}
\end{equation}
In fact, for the given $\Delta,{\cal W}$, the critical points $\sigma_c^i$ 
satisfy
$$
\sqrt{1+\alpha -{v(\sigma_c^i)\over M}}={\sqrt{1-{\bar M\over M}}\over \Delta}
\geq {\sqrt{1-{\bar M\over M}}\over \Delta_{Max}} 
$$
or equivalently
$$
{v(\sigma_c^i)\over M}\leq 1+\alpha -{1-{\bar M\over M}\over \Delta^2_{Max}}~.
$$
Choosing $\mu$ to satisfy
\begin{equation}
{v(\mu)\over M}>  1 -{7\over 8}\ {1-{\bar M\over M}\over \Delta^2_{Max}} 
\label{ineq1}
\end{equation}
and using (\ref{delmax}), we obtain
$$
{v(\sigma_c^i)\over M}\leq 
 1+\alpha -{1-{\bar M\over M}\over \Delta^2_{Max}}\leq
 1 -{7\over 8}\ {1-{\bar M\over M}\over \Delta^2_{Max}} < {v(\mu)\over M}
$$
and therefore we have $\sigma^i_c\in (\mu,2\pi-\mu)$. Then, for 
all $\sigma\in [0,\mu]$, we have
$$
\norm{\Phi'(\sigma)}\geq {\sqrt{1-{\bar M\over M}}
\over \sqrt{1+\alpha-{v(\sigma)\over M}}}-\Delta
\geq {\sqrt{1-{\bar M\over M}}
\over \sqrt{1+\alpha-{v(\mu)\over M}}}-\Delta_{Max}
\geq {\sqrt{1-{\bar M\over M}}
\over 16\sqrt{1+\alpha-{v(\mu)\over M}}}
$$
as soon as
\begin{equation}
\Delta_{Max} \leq {15\over 16}\ {\sqrt{1-{\bar M\over M}}
\over \sqrt{1+\alpha-{v(\mu)\over M}}}~,
\label{ineq2}
\end{equation}
which is satisfied if
$$
{v(\mu)\over M}\geq 1+\alpha - \left ( {15\over 16}\right )^2
{1-{\bar M\over M}\over \Delta^2_{Max}} .
$$
From (\ref{delmax}) and (\ref{ineq1}) we obtain
$$
1+\alpha - \left ( {15\over 16}\right )^2{1-{\bar M\over M}\over \Delta^2_{Max}}
\leq 1 - {7 \over 8}\ {1-{\bar M\over M}\over \Delta^2_{Max}} \leq {v(\mu)\over M}~.
$$
Therefore, for $\mu$ satisfying (\ref{ineq1}), for all
$\sigma\in [0,\mu]$  we have
$$
\norm{\Phi'(\sigma)}\geq {\Delta_{Max}\over 15}  .
$$
Analogously, we have
$$
\norm{\Phi'(\sigma)s_\alpha(\sigma)} \geq \sqrt{2M\over \norm{A}}
\sqrt{1-{\bar M\over M}}\left (1 - \Delta {\sqrt{1+\alpha-{v(\sigma)\over M}}
\over \sqrt{1-{\bar M\over M}}} \right  ) \geq
$$
$$
\geq  \sqrt{2M\over \norm{A}}\sqrt{1-{\bar M\over M}}\left (1 - \Delta_{Max} {\sqrt{1+\alpha-{v(\mu)\over M}}\over \sqrt{1-{\bar M\over M}}} \right  )\geq
{1\over 16}\sqrt{M\over \norm{A}}\sqrt{1-{\bar M\over M}} .
$$
Let us now consider an infinitely
differentiable function $\rho(x,\mu)$ such that 
$\rho(x,\mu)=0$ for $x\leq \mu/2$ and $x\geq 2\pi-\mu/2$, 
$\rho(x,\mu)=1$ for $x\in [\mu,2\pi-\mu]$, 
$ {\partial^j\over \partial x^j}\rho(\mu/2,\mu)=
{\partial^j\over \partial x^j}\rho(2\pi-\mu/2,\mu)$ for all $j\geq 0$. 
Then, we define
$$
\eta(\sigma)=\rho(\sigma;\mu)[s_\alpha(\sigma)]^{m-1}
$$
and
$$
\hat I(\norm{\cal W},\Delta )=\int_{0}^{2\pi}
\eta(\sigma)e^{i\norm{\cal W}\Phi(\sigma)}d\sigma .
$$
The integrand of $\hat I(\norm{\cal W},\Delta )$ is smooth and bounded
for all $m\geq 0$, and vanishes at the extrema together with all its
derivatives.  Therefore it has the form suitable for the application
of the rigorous version of PSP (see, for example, \cite{BleHan}). As a
consequence, taking into account that the phase $\Phi(\sigma)$ has two
non-degenerate critical points $\sigma_c^1,\sigma_c^2$, $\hat I
(\norm{\cal W},\Delta )$ is represented by (\ref{rigorous1}) with
$a_1,a_2$ satisfying the limits (\ref{limits1}). It remains to
estimate the integral
$$
I(\norm{W},\Delta)-\hat I(\norm{W},\Delta)
= \int_{0}^{2\pi}(1-\rho(\sigma;\mu))[s_\alpha(\sigma )]^{m-1}e^{i\norm{\cal W}\Phi(\sigma)}d\sigma 
$$
$$
=\int_{0}^{\mu}(1-\rho(\sigma;\mu))[s_\alpha(\sigma )]^{m-1}e^{i\norm{\cal W}\Phi(\sigma)}d\sigma+
\int_{2\pi -\mu}^{2\pi}(1-\rho(\sigma;\mu))[s_\alpha(\sigma )]^{m-1}e^{i\norm{\cal W}\Phi(\sigma)}d\sigma .
$$
We prove that there exists a constant $\kappa$ independent on
$\norm{\cal W}$, $\Delta$ and $m$, such that
\begin{equation}
\norm{I(\norm{W},\Delta)-\hat I(\norm{W},\Delta)}\leq \Norm{s_\alpha}^{m-1}
{\kappa \over \norm{\cal W}}  ,
\label{diffintegrals}
\end{equation}
and therefore also the integral $I(\norm{W},\Delta)$ has the representation 
(\ref{rigorous1}) with $a_1,a_2$ satisfying the limits (\ref{limits1}).

Since the phase $\Phi(\sigma)$ 
has no stationary points in $[0,\mu]$, and if $m\geq 1$, 
integrating by parts we obtain
\begin{equation}
\int_{0}^{\mu}(1-\rho(\sigma;\mu))[s_\alpha(\sigma )]^{m-1}e^{i\norm{\cal W}\Phi(\sigma)}d\sigma= -{[s_\alpha(0)]^{m-1}e^{i\theta(0)}\over i \norm{\cal W}\Phi'(0)}-
\int_{0}^{\mu}\left ({(1-\rho(\sigma;\mu))[s_\alpha(\sigma)]^{m-1}
\over  i\norm{\cal W}\Phi'(\sigma)}\right )'
e^{i\norm{\cal W}\Phi(\sigma)}d\sigma .
\label{integraleperpartinondeg}
\end{equation}
We consider the following cases: 
\begin{itemize}
\item[-] {\it If $m\geq 3$ or $m=1$,} using (\ref{phiprimo})
we obtain
$$
\norm{{[s_\alpha(0)]^{m-1}e^{i\theta(0)}\over i\norm{\cal W}\Phi'(0)}}\leq 
\norm{[s_\alpha(0)]}^{m-1}{k_2 \over \norm{\cal W}}  .
$$
Then we estimate
$$
\norm{\int_{0}^{\mu}\left ({(1-\rho(\sigma;\mu)) s^{m-1}_\alpha(\sigma)
\over  i\norm{\cal W}\Phi'(\sigma)}\right )'
e^{i\norm{\cal W}\Phi(\sigma)}d\sigma} 
\leq \int_{0}^{\mu}
\norm{\rho'(\sigma;\mu) s_\alpha^{m-1}(\sigma)
\over  i\norm{\cal W}\Phi'(\sigma)}d\sigma 
$$
\begin{equation}
+\int_{0}^{\mu}\norm{(1-\rho(\sigma;\mu))(m-1)s^{m-3}_\alpha(\sigma) 
v'(\sigma)
\over  iA\norm{\cal W}\Phi'(\sigma)}d\sigma+
\int_{0}^{\mu}\norm{(1-\rho(\sigma;\mu)) s^{m-1}_\alpha(\sigma)}
\norm{ \left ( {1\over  \norm{\cal W}\Phi'(\sigma))}\right )'} d\sigma
~.
\label{threetermsnondeg}
\end{equation}
Using again (\ref{phiprimo}) we obtain
$$
\int_{0}^{\mu}
\norm{\rho'(\sigma;\mu) s_\alpha^{m-1}(\sigma)
\over  i\norm{\cal W}\Phi'(\sigma)}d\sigma+
\int_0^\mu\norm{M(1-\rho(\sigma;\mu))(m-1)s^{m-3}_\alpha(\sigma) 
\over  iA\norm{\cal W}\Phi'(\sigma)}d\sigma \leq 
\Norm{s_\alpha}^{m-1}
{k_{3}\over \norm{\cal W}}~.
$$
Since $1/\Phi'(\sigma)$ is strictly
monotone in $[0,\mu]$, its derivative has the same sign in $[0,\mu]$, 
and therefore we have
$$
\int_0^\mu \norm{(1-\rho(\sigma;\mu)) s^{m-1}_\alpha(\sigma)}
\norm{ \left ( {1\over  \norm{\cal W}\Phi'(\sigma))}\right )'} 
d\sigma \leq \Norm{ s_\alpha}^{m-1}
\int_0^\mu \norm{ \left ( {1\over  \norm{\cal W}\Phi'(\sigma))}\right )'} 
d\sigma =
$$
$$
=\Norm{ s_\alpha}^{m-1}\norm{
\int_0^\mu \left ( {1\over  \norm{\cal W}\Phi'(\sigma))}\right )' 
d\sigma} =  \Norm{ s_\alpha}^{m-1}
\norm{ {1\over   \norm{\cal W}\Phi'(\mu)}- 
{1\over   \norm{\cal W}\Phi'(0 )}} \leq
\Norm{ s_\alpha}^{m-1}{k_{4}\over \norm{\cal W}} ~.
$$
By collecting all these estimates, we obtain
$$
\norm{\int_{0}^{\mu}(1-\rho(\sigma;\mu))[s_\alpha(\sigma )]^{m-1}e^{i\norm{\cal W}\Phi(\sigma)}d\sigma}\leq \Norm{ s_\alpha}^{m-1}{k_{5}\over \norm{\cal W}}
$$
with $k_{5}>0$ independent on $\norm{\cal W}$, $\Delta$ and $m$. 

\item[-] {\it If $m=2$,} we estimate the border contribution in 
(\ref{integraleperpartinondeg}), and first and the 
third integral in (\ref{threetermsnondeg}) as in the case $m\geq 3$. It 
remains to estimate the second integral, 
whose denominator is only apparently divergent, since because
of (\ref{phiprimo}) we have 
$\norm{\Phi'(\sigma)s_\alpha(\sigma)}\geq k_1$. Therefore we have
$$
\int_{0}^{\mu}\norm{(1-\rho(\sigma;\mu))
\over  i\norm{A}\norm{\cal W}\Phi'(\sigma)s_\alpha(\sigma) } d\sigma \leq 
{k_{6}\over \norm{\cal W}}~.
$$

\item[-]  {\it If $m=0$,} for any arbitrary small $\xi\in (0,\mu)$, 
we have
$$
\int_{\xi}^{\mu}(1-\rho(\sigma;\mu))[s_\alpha(\sigma )]^{-1}e^{i\norm{\cal W}\Phi(\sigma)}d\sigma
= -{1-\rho(\xi;\mu)\over i \norm{\cal W}\Phi'(\xi)s_\alpha(\xi)}e^{i\norm{\cal W}\Phi(\xi)}+
$$
$$
+\int_{\xi}^{\mu} {\rho'(\sigma;\mu)\over i \norm{\cal W}\Phi'(\sigma)
s_\alpha(\sigma)}e^{i\norm{\cal W}\Phi(\sigma)}d\sigma-
\int_{\xi}^{\mu} (1-\rho(\sigma;\mu))
\left ( {1\over i \norm{\cal W}\Phi'(\sigma)
s_\alpha(\sigma)}\right )'e^{i\norm{\cal W}\Phi(\sigma)}d\sigma .
$$
Using (\ref{phiprimo}) we obtain, uniformly on $\xi$,
$$
\norm{ {1-\rho(\xi;\mu)\over i \norm{\cal W}\Phi'(\xi)s_\alpha(\xi)}e^{i\norm{\cal W}\Phi(\xi)}+
\int_{\xi}^{\mu} {\rho'(\sigma;\mu)\over i \norm{\cal W}\Phi'(\sigma)
s_\alpha(\sigma)}e^{i\norm{\cal W}\Phi(\sigma)}d\sigma} \leq {k_7\over \norm{\cal W}} ~.
$$
Since $1/(\Phi'(\sigma)s_\alpha(\sigma))$, is 
strictly monotone in $[0,\mu]$, by proceeding as in 
the cases $m\geq 1$, we obtain
$$
\norm{\int_{\xi}^{\mu}(1-\rho(\sigma;\mu))[s_\alpha(\sigma )]^{-1}e^{i\norm{\cal W}\Phi(\sigma)}d\sigma}\leq {k_8\over \norm{\cal W}}
$$
uniformly in $\xi$, so that
$$
\norm{\int_{0}^{\mu}(1-\rho(\sigma;\mu))[s_\alpha(\sigma )]^{-1}e^{i\norm{\cal W}\Phi(\sigma)}d\sigma}\leq {k_8\over \norm{\cal W}}  .
$$

\end{itemize}
\vskip 0,4 cm By repeating the argument to estimate the integral on
$[2\pi-\mu,2\pi]$ we obtain (\ref{diffintegrals}).  \hfill $\square$

\vskip 0.4 cm
Lemma 1 provides an explicit criterion allowing to classify
Melnikov integrals as belonging to the categories (I) or (II)
depending on the values of of ${\cal N},{\cal W}$. Precisely, if
${\cal N}\cdot {\cal W}>0$ the phase is considered in the category
(I), and the contribution of the corresponding Melnikov integral to
the Arnold diffusion will be considered negligible. Instead, if ${\cal
  N}\cdot{\cal W}<0$, and $|{\cal N}|/|{\cal W}|$ satisfies strictly
(\ref{condNWalpha}), the phase has two non-degenerate critical points.
Then we distinguish two subcases:
\begin{itemize}
\item[] - $|{\cal N}|/|{\cal W}|$ is not too close to its lower
  extremum, according to to a criterion specified by Lemma 2
  below. Then, the phase is considered in the category (II) and the
  contribution of the corresponding Melnikov integral to the Arnold
  diffusion can be estimated analytically (\ref{rigorous1}).
\item[] - ${\cal N}\cdot {\cal W}<0$ and ${|{\cal N}|\over |{\cal
    W}|}$ suitably close to its lower extremum. We find that such
  a term, while formally 'stationary', contributes to the Melnikov
  integral similarly as 'quasi-stationary' terms satisfying
  \begin{equation}
    {|{\cal N}|\over |{\cal W}|}< 
    {\sqrt{1-{\bar M\over M}} \over \sqrt{1+\alpha -{\bar M\over M}}} = Q_{\alpha} .
    \label{underdeg}
  \end{equation}
  In fact, a careful investigation of the transition of ${|{\cal
      N}|\over |{\cal W}|}$ from values higher than $Q_{\alpha}$ to
  smaller ones, reveals that the transition corresponds to a
  degenerate critical point for the phase.  The corresponding integral
  blows to values of order $1/\norm{\cal W}^{1\over 3}$, at the
  transition, and depending linearly on the distance $\delta =
    \frac{{\cal N}}{{\cal W}} - Q_{\alpha}$, for small
    $|\delta|$. Thus, these intermediate cases will be considered in
  the quasi-stationary category (III).

\end{itemize}
  The values of the  Melnikov integrals for terms in category (III)
  are estimated according to the following:
\vskip 0.2 cm
 \noindent
{\bf Lemma 2.} {\it  Let the potential $v(\sigma)$ have only one local 
non--degenerate maximum at $\sigma=0$ and one local non--degenerate minimum 
at $\sigma=\bar \sigma$. Let us consider $\epsilon$ suitably 
small and $\alpha \in (0,\Norm{r^N})$. For any  
phase $\theta(\sigma )$ defined by the labels $m,\nu,k$
such that ${\cal N}>0$, ${\cal W}<0$ and
\begin{equation}
\norm{\cal N}=\norm{\cal W}\left ({\sqrt{1-{\bar M\over M}} \over \sqrt{1+\alpha -{\bar M\over M}}} -\delta \right)  
\label{NWdelta}
\end{equation}
with some $\delta > 0$, and defining
\begin{equation}
I(\norm{\cal W},\delta )=\int_{0}^{2\pi}[s_\alpha(\sigma)]^{m-1}e^{i\theta(\sigma)}d\sigma ,
\label{qsdetune0}
\end{equation}
we have
\begin{equation}
I(\norm{\cal W},\delta )= \hat I(\norm{\cal W},\delta )+
{b(\norm{\cal W},\delta )\over \norm{\cal W}}
\end{equation}
where
\begin{eqnarray}
&\left ({\partial^j \over \partial \delta^j}\hat I(\norm{\cal W},\delta )\right )_{\Big \vert \delta=0} =e^{i \theta_*} c_j\norm{\cal W}^{2j-1\over 3}a_j(\norm{\cal W})& ,
\ \ j\geq 0\label{asym1}\\
\cr
&\lim_{\norm{\cal W}\rightarrow +\infty} a_j(\norm{\cal W}) = 1  ,\ \ j\geq 0&  \label{asym2} \\ 
\cr
&\norm{b(\norm{\cal W},\delta )} \leq \kappa& 
\end{eqnarray}
with constants $\kappa$ and $c_j$ independent on 
$\norm{W}$ and $\delta$, and $\theta_*=\theta(\bar \sigma)_{\vert \delta=0}$.
In particular, we have
\begin{equation}
c_0={\sqrt{3}\Gamma(4/3)
\Big ({2M\over \norm{A}} (1+\alpha-{\bar M\over M})\Big )^{m-1\over 2}\over
\left ( {\sqrt{1-{\bar M\over M}}\over 12M\Big (1+\alpha-{\bar M\over M}
\Big )^{3\over2}}v''(\bar \sigma)\right )^{1\over 3}}
\ \ ,\ \ c_1=-{{\Gamma(2/3)\over \sqrt{3}}
\Big ({2M\over \norm{A}} (1+\alpha-{\bar M\over M})\Big )^{m-1\over 2}\over
\left ( {\sqrt{1-{\bar M\over M}}\over 12M\Big (1+\alpha-{\bar M\over M}
\Big )^{3\over2}}v''(\bar \sigma)\right )^{2\over 3}}  ~, \label{asym3}
\end{equation}
and, for all $j\geq 2$,
\begin{equation}
\norm{c_j}= {{\Gamma((j+1)/3)\over 3}
\Big ({2M\over \norm{A}} (1+\alpha-{\bar M\over M})\Big )^{m-1\over 2}\over
\left ( {\sqrt{1-{\bar M\over M}}\over 12M\Big (1+\alpha-{\bar M\over M}
\Big )^{3\over2}}v''(\bar \sigma)\right )^{j+1\over 3}}\alpha_j  \label{asymj}
\end{equation}
with $\alpha_j \in \{0,1,\sqrt{3},2\}$ depending on $j$. 
}
\vskip 0.2 cm
\noindent
{\bf Proof of Lemma 2.} Let us choose $\mu>0$ small 
enough, but independent on $\norm{\cal W}$, $\delta$ and $\alpha$; 
let us define an infinitely
differentiable function $\rho(x,\mu)$ such that 
$\rho(x,\mu)=0$ for $x\leq 0$ and $x\geq 2\pi$; 
$\rho(x,\mu)=1$ for $x\in [\mu,2\pi-\mu]$; 
$ {\partial^j\over \partial x^j}\rho(0,\mu)=
{\partial^j\over \partial x^j}\rho(2\pi,\mu)$ for all $j\geq 0$. 
Then, we define
$$
\eta(\sigma)=\rho(\sigma;\mu)[s_\alpha(\sigma)]^{m-1}
$$
and
$$
 \hat I(\norm{\cal W},\delta )=\int_{0}^{2\pi}
\eta(\sigma)e^{i\theta(\sigma)}d\sigma .
$$
Let us preliminarly write
the phase $\theta(\sigma)$ and its derivative by replacing ${\cal N}$ 
using (\ref{NWdelta}):
\begin{eqnarray}
\theta &=& \theta(\bar \sigma)+\norm{\cal W}\left [\left ( {\sqrt{1-{\bar M\over M}} \over 
\sqrt{1+\alpha -{\bar M\over M}}} -\delta \right)(\sigma -\bar \sigma)
-{ \sqrt{1-{\bar M\over M}}} 
\int_{\bar \sigma}^\sigma {dx\over  \sqrt{1+\alpha - {v(\sigma ) \over M}}}
\right ]
\hskip 0.2 cm\label{theta0}\\
 \theta' &=& -\norm{\cal W}\left [\delta+\sqrt{1-{\bar M\over M}}\left (
{1\over \sqrt{1+\alpha -{v(\sigma )\over M}}}-
{1\over \sqrt{1+\alpha -{\bar M\over M}}}\right )\right ]
\label{theta1}
\end{eqnarray}
as well as the expansions at $\sigma=\bar \sigma$
\begin{equation}
\theta = \theta(\bar \sigma)-\norm{\cal W}\left [
\delta (\sigma -\bar \sigma) 
+{\sqrt{1-{\bar M\over M}}\over 12 M}{v''(\bar \sigma)\over  
\left (1+\alpha -{\bar M\over M}\right )^{3\over 2}}(\sigma -\bar \sigma)^3
+{\cal O}(\sigma-\bar\sigma)^4\right ]\label{taytheta0}
\end{equation}
In the following we denote by $k_1,k_2,\ldots $ suitable 
constants which do not depend on $m,{\cal W},\delta,\epsilon,\alpha$, 
while they may depend on $\mu$. Since (for small $\mu$)  
$\bar \sigma \in (\mu, 2\pi-\mu)$, for all 
$\sigma\in [0,\mu]$ we have (see (\ref{theta1}))
\begin{equation}
\norm{\theta'(\sigma)} \geq \norm{\cal W}\sqrt{1-{\bar M\over M}}\left (
{1\over \sqrt{1+\alpha -{v(\sigma )\over M}}}-
{1\over \sqrt{1+\alpha -{\bar M\over M}}}\right )\geq
\norm{\cal W} k_1 ~,
\label{thetaintervallo}
\end{equation}
as well as
\begin{equation}
\norm{\theta'(\sigma)s_\alpha (\sigma )}\geq  
\norm{\cal W}\sqrt{1-{\bar M\over M}}\sqrt{2 M\over \norm{A}}\left (
1- {\sqrt{1+\alpha -{v(\sigma )\over M}}\over  
\sqrt{1+\alpha -{\bar M\over M}}}\right )\geq
\norm{\cal W} k_1 ~.
\label{thetasintervallo}
\end{equation}
To estimate
$$
{b(\norm{\cal W},\delta)\over \norm{\cal W}}=
I(\norm{W},\delta)-\hat I(\norm{W},\delta)
= \int_{0}^{2\pi}(1-\rho(\sigma;\mu))[s_\alpha(\sigma )]^{m-1}e^{i\theta(\sigma)}d\sigma 
$$
$$
=\int_{0}^{\mu}(1-\rho(\sigma;\mu))[s_\alpha(\sigma )]^{m-1}e^{i\theta(\sigma)}d\sigma+
\int_{2\pi -\mu}^{2\pi}(1-\rho(\sigma;\mu))[s_\alpha(\sigma )]^{m-1}e^{i\theta(\sigma)}d\sigma ~,
$$
we first notice that the phase $\theta(\sigma)$ 
has no stationary points in $[0,\mu]$ and therefore integrating 
by parts we obtain
\begin{equation}
\int_{0}^{\mu}(1-\rho(\sigma;\mu))[s_\alpha(\sigma )]^{m-1}e^{i\theta(\sigma)}d\sigma= 
-{[s_\alpha(0)]^{m-1}e^{i\theta(0)}\over i\theta'(0)}-
\int_{0}^{\mu}\left ({(1-\rho(\sigma;\mu))[s_\alpha(\sigma)]^{m-1}
\over  i\theta'(\sigma)}\right )'
e^{i\theta(\sigma)}d\sigma ~.
\label{integraleperparti}
\end{equation}
We consider the following cases:
\begin{itemize}
\item[-] {\it If $m\geq 3$ or $m=1$,} using (\ref{thetaintervallo})
we obtain
$$
\norm{{[s_\alpha(0)]^{m-1}e^{i\theta(0)}\over i\theta'(0)}}\leq 
\norm{[s_\alpha(0)]}^{m-1}{k_2\over \norm{\cal W}} .
$$
Then we can estimate
$$
\norm{\int_{0}^{\mu}\left ({(1-\rho(\sigma;\mu)) s^{m-1}_\alpha(\sigma)
\over  i\theta'(\sigma)}\right )'
e^{i\theta(\sigma)}d\sigma} \leq \int_{0}^{\mu}
\norm{\rho'(\sigma;\mu) s_\alpha^{m-1}(\sigma)
\over  i\theta'(\sigma)}d\sigma+
$$
\begin{equation}
+\int_{0}^{\mu}\norm{(1-\rho(\sigma;\mu))(m-1)s^{m-3}_\alpha(\sigma) 
v'(\sigma)
\over  iA\theta'(\sigma)}d\sigma+
\int_{0}^{\mu}\norm{(1-\rho(\sigma;\mu)) s^{m-1}_\alpha(\sigma)}
\norm{ \left ( {1\over  \theta'(\sigma))}\right )'} d\sigma ~.
\label{threeterms}
\end{equation}
Using (\ref{thetaintervallo}) we obtain
$$
\int_{0}^{\mu}
\left ( \norm{\rho'(\sigma;\mu) s_\alpha^{m-1}(\sigma)
\over  i\theta'(\sigma)}+\norm{M(1-\rho(\sigma;\mu))(m-1)s^{m-3}_\alpha(\sigma) 
\over  iA\theta'(\sigma)}\right ) d\sigma \leq 
\Norm{s_\alpha}^{m-1}
{k_3\over \norm{\cal W}} ~.
$$
Since $1/\theta'(\sigma)$ is strictly
monotone in $[0,\mu]$, its derivative has the same sign in $[0,\mu]$, 
and therefore we have
$$
\int_0^\mu \norm{(1-\rho(\sigma;\mu)) s^{m-1}_\alpha(\sigma)}
\norm{ \left ( {1\over  \theta'(\sigma))}\right )'} 
d\sigma \leq \Norm{ s_\alpha}^{m-1}
\int_0^\mu \norm{ \left ( {1\over  \theta'(\sigma))}\right )'} 
d\sigma =
$$
$$
=\Norm{ s_\alpha}^{m-1}\norm{
\int_0^\mu \left ( {1\over  \theta'(\sigma))}\right )' 
d\sigma} =  \Norm{ s_\alpha}^{m-1}
\norm{ {1\over  \theta'(\mu)}- {1\over  \theta'(0 )}} \leq
 \Norm{ s_\alpha}^{m-1}{k_4\over \norm{\cal W}}  ~.
$$
By collecting all these estimates, we obtain
$$
\norm{\int_{0}^{\mu}(1-\rho(\sigma;\mu))[s_\alpha(\sigma )]^{m-1}e^{i\theta(\sigma)}d\sigma}\leq \Norm{ s_\alpha}^{m-1}{k_5\over \norm{\cal W}} ~.
$$

\item[-] {\it If $m=2$,} we estimate the border contribution in 
(\ref{integraleperparti}), and first and the 
third integral in (\ref{threeterms}) as in the case $m\geq 3$. It 
remains to estimate the second integral, 
whose denominator is only apparently divergent, since because
of (\ref{thetasintervallo}) we have 
$\norm{\theta'(\sigma)s_\alpha(\sigma)}\geq \norm{\cal W}k_1$. Therefore we have
$$
\int_{0}^{\mu}\norm{(1-\rho(\sigma;\mu))
\over  i\norm{A}\theta'(\sigma)s_\alpha(\sigma) } d\sigma \leq 
{k_{6}\over \norm{\cal W}} .
$$
 
\item[-]  {\it If $m=0$,} for any arbitrary small $\xi\in (0,\mu)$, 
we have
$$
\int_{\xi}^{\mu}(1-\rho(\sigma;\mu))[s_\alpha(\sigma )]^{-1}e^{i\theta(\sigma)}d\sigma
= -{1-\rho(\xi;\mu)\over i \theta'(\xi)s_\alpha(\xi)}e^{i\theta(\xi)}+
$$
$$
+\int_{\xi}^{\mu} {\rho'(\sigma;\mu)\over i \theta'(\sigma)
s_\alpha(\sigma)}e^{i\theta(\sigma)}d\sigma-
\int_{\xi}^{\mu} (1-\rho(\sigma;\mu))
\left ( {1\over i \theta'(\sigma)
s_\alpha(\sigma)}\right )'e^{i\theta(\sigma)}d\sigma .
$$
Using (\ref{thetasintervallo}) we obtain uniformly on $\xi$
$$
\norm{ {1-\rho(\xi;\mu)\over i \theta'(\xi)s_\alpha(\xi)}e^{i\theta(\xi)}+
\int_{\xi}^{\mu} {\rho'(\sigma;\mu)\over i \theta'(\sigma)
s_\alpha(\sigma)}e^{i\theta(\sigma)}d\sigma} \leq {k_7\over \norm{\cal W}}  .
$$
Since $1/(\Phi'(\sigma)s_\alpha(\sigma))$, is 
strictly monotone in $[0,\mu]$, by proceeding as in 
the cases $m\geq 1$, we obtain
$$
\norm{\int_{\xi}^{\mu}(1-\rho(\sigma;\mu))[s_\alpha(\sigma )]^{-1}e^{i\theta(\sigma)}d\sigma}\leq {k_8\over \norm{\cal W}}~,
$$
uniformly in $\xi$, so that
$$
\norm{\int_{0}^{\mu}(1-\rho(\sigma;\mu))[s_\alpha(\sigma )]^{-1}e^{i\theta(\sigma)}d\sigma}\leq {k_8\over \norm{\cal W}}  ~.
$$

\end{itemize}
\vskip 0,4 cm
By repeating the argument to estimate the integral on $[2\pi-\mu,2\pi]$
we obtain that there exists a constant $\kappa$ independent on 
$\norm{\cal W}$ and $\delta$ such 
that $\norm{b(\norm{\cal W},\delta)}\leq \kappa$. 

Then, since the integral $\hat I(\norm{\cal W},\delta)$ is smooth 
with respect to $\delta$ at $\delta=0$, we have
$$
\left ({\partial^j \over \partial \delta^j}\hat I(\norm{\cal W},\delta )
\right )_{\Big \vert \delta=0}=(-i)^j\norm{\cal W}^{j}e^{i\theta_*}
\int_0^{2\pi}\eta(\sigma)(\sigma-\bar \sigma)^je^{i\norm{\cal W}\Phi(\sigma)}d\sigma 
$$
where
$$
\Phi(\sigma)=\left [{\sqrt{1-{\bar M\over M}} \over 
\sqrt{1+\alpha -{\bar M\over M}}}(\sigma -\bar \sigma)
-{ \sqrt{1-{\bar M\over M}}} 
\int_{\bar \sigma}^\sigma {dx\over  \sqrt{1+\alpha - {v(\sigma ) \over M}}}
\right ] .
$$
Therefore, by using the degenerate version of the principle of 
stationary-phase (see \cite{Erdelyi}), by identifying in $\norm{\cal W}$ the 
large parameter, and 
by considering that $\eta(\sigma)(\sigma -\bar\sigma)^j$ vanishes
with all its derivatives at $\sigma=0,2\pi$, we obtain (\ref{asym1}),
 (\ref{asym2}), (\ref{asym3}), (\ref{asymj}).
\hfill $\square$
\vskip 0,4 cm Lemma 2 allows us to study the transition in the
representations of the Melnikov integrals from the regime of
stationary phases to the regime of non stationary phases. Hence:

\begin{itemize}
\item[- ]{\it In the case $\delta \geq 0$,} a non-stationary phase is 
considered quasi-stationary, and the corresponding Melnikov integral is approximated by
\begin{equation}
I(\norm{\cal W},\delta )= c_0{e^{i\theta_*}\over 
\norm{\cal W}^{1\over 3}}-\norm{c_1}\delta e^{i\theta_*}\norm{\cal W}^{1\over 3} +\ldots
\label{linearI}
\end{equation}
if
\begin{equation}
0\leq \delta \leq \delta_c={\norm{c_1}\over c_0}{1\over  \norm{W}^{2/3}}  ,
\label{deltac}
\end{equation}
otherwise the phase is considered non-stationary and the 
corresponding Melnikov integral is neglected. 

\item[- ] {\it In the case $\delta<0$,} since $I(\norm{\cal
  W},\delta)$ is smooth in $\delta$, at $\delta=0$, we compare two
  estimates: one coming from Lemma 1 (stationary phase approximation)
  and another coming from the extension of the linear
  law~\eqref{linearI} (quasi--stationary phase approximation) to small
  negative $\delta$. We find that, for negative $\delta$ suitably
  close to $0$ the quasi-stationary phase approximation provides a
  better estimate with respect to the stationary phase approximation.
  To determine a threshold to decide which one to use, we compared the
  numerical computation of the integrals with the estimates provided
  by both Lemmas. Let us, for example, consider $v(\sigma
  )=\cos\sigma$; for $\alpha=0$, we have
\begin{equation}
\theta(\sigma)= \theta_0+{\cal N}\sigma +2{\cal W}\ln\tan(\sigma/4)~,
\label{thsigpend}
\end{equation}
and the integrals
\begin{equation}
\Delta{\cal I} = \int_0^{2\pi} 
\cos\left({\cal N}(\sigma-\pi) + 2{\cal W}\ln\tan(\sigma/4)\right)d\sigma ~.
\label{inte} 
\end{equation}
\end{itemize}

\begin{figure*}
\includegraphics[width=.99\textwidth]{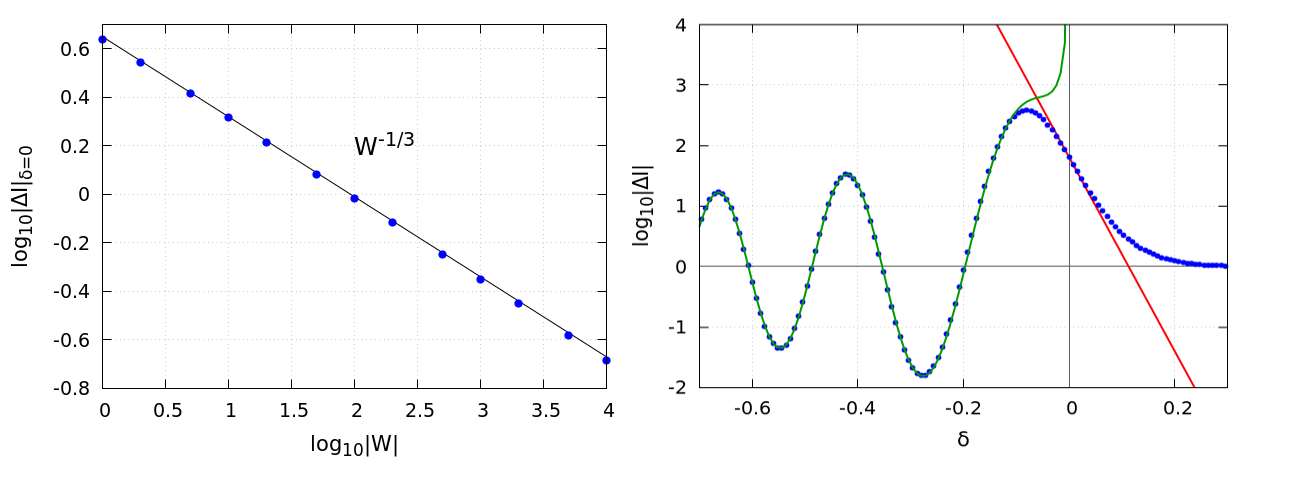}
\caption{\small In the left panel we computed numerically the values
  of $\Delta {\cal I}$ defined in (\ref{inte}) for $\delta =0$ and
  several values of $|{\cal W}|$ (blue points).  The blue line
  corresponds to the asymptotic law $1/\norm{\cal W}^{1\over 3}$, that
  well approximate the values of the integrals for the whole interval
  of $|{\cal W}|$ considered. In the right panel we compare the
  numerical values of the integrals $\Delta {\cal I}$ (the blue dots),
  with the corresponding estimate provided by the stationary-phase
  approximation (green curve) and by the linear law
  (\ref{linearI}) (red line), for the sample value $\norm{W}=15$ and
  several values of $\delta$.}
\label{fig:quasi}
\end{figure*}

Figure~\ref{fig:quasi} shows the values of the integrals $\Delta{\cal
  I}$, computed numerically, for several values of ${\cal W}<0$, and
fixed $\delta$ (left panel), or for fixed ${\cal W}$ and several
values of $\delta$ (right panel). The left panel shows that the value
of the integrals as computed numerically by solving Eq.~(\ref{inte})
(blue dots) is well approximated by the corresponding asymptotic law
$1/\norm{\cal W}^{1\over 3}$ for the values of $|{\cal W}|$ considered
(blue line). In the right-panel we compare the numerical computations
of (\ref{inte}) (blue dots), with the corresponding estimate provided
by the stationary-phase approximation (green curve) and with the
linear law (\ref{linearI}) (red line), for the sample value
$\norm{W}=15$ (very similar pictures are obtained for different
values).  We see that the stationary phase estimates reproduce well
the values of the integrals for $\delta \leq -\delta_c/2$.  For
$\delta \in [-\delta_c/2,0]$ we have a divergence of the stationary
phase approximation formula, indicating that the approximation is no
more valid since we are entering the regime of quasi-stationary
phase. In fact, we observe that the linear law (\ref{linearI})
represents much better the value of the integral for both positive and
negative $\delta$ in the interval $-\delta_c/2 < \delta < \delta_c$,
and therefore, we use Eq.~\eqref{linearI} with $c_0$, $c_1$ given by
Lemma 2 also to estimate those integrals. By using the formula down to
$\delta=\delta_c$ we introduce some errors, which could be reduced by
considering the non linear corrections (see formula (\ref{nonlinearI})
below). On the other hand, $\delta_c$ as computed from (\ref{linearI})
(represented by the point at which the linear law crosses the x--axis)
is clearly underestimated, since the non-linear contributions
determine that $\Delta{\cal I}$ has a tail extending only
asymptotically to zero (see remark (xiii)).
\vskip 0.2 cm
\noindent
{\bf Remarks:} 
\vskip 0.2 cm
\noindent
\begin{itemize}

\item[(xiii)]
  The non-linear terms of the expansion (\ref{linearI})
provide corrections to the critical value $\delta_c$ necessary
to discriminate between quasi-stationary or non stationary phase. 
Since $c_2=0$ and $c_3<0$, a more careful analysis of the non-linear terms  
provides
 \begin{equation}
I(\norm{\cal W},\delta )\sim e^{i\theta_*} \left ({c_0\over 
\norm{\cal W}^{1\over 3}}-\norm{c_1}\delta \norm{\cal W}^{1\over 3}
+b \delta^2 \norm{\cal W}^{2\over 3}-\norm{c_3} \delta^3  \norm{\cal W}^{5\over 3}
+\ldots \right )
\label{nonlinearI}
\end{equation}
so that, for smaller values of $\norm{\cal W}$ the quadratic and the cubic 
terms in $\delta$ can produce a small variation of $\delta_c$.

\item[(xiv)]
  The above analysis applies to the terms with suitably
large values of
$$
\norm{\cal W}={\norm{\Omega}\over \sqrt{2 \norm{A}\epsilon (M-\bar M)}}=
{\norm{k\cdot \omega_*}\over \sqrt{2 \norm{A}\epsilon (M-\bar M)}} ,
$$ 
where $\norm{k\cdot \omega_*}$ are non-resonant divisors. In 
Nekhoroshev theorem, also in the more generic steep case, 
if $\norm{k}\leq N$ these divisors 
are estimated by (see for example, the discussion in
the introduction of \cite{GCB16})
$$
\norm{k\cdot \omega_*} \geq N\sqrt{\epsilon}
$$
so that, correspondingly, $\norm{\cal W}$ is large. For
$\norm{k}\geq N$, we have no theoretical lower bounds on $\norm{W}$.
However, by analyticity, the corresponding Fourier harmonics in the
original perturbation are exponentially small in $\norm{k}$, and hence
also exponentially small in $N$.  Thus, after the normalization
procedure we do not expect to find such harmonics in the dominant
terms of $r^N$.

\item[(xv)] Lemma 1 and 2 are derived by considering the upper branch
  of the separatrix solution $\theta(\sigma)$ and $\alpha> 0$. 
    Equivalent results are found for the lower branch, and for
  $\alpha<0$, after some obvious modifications in the formulas.

\end{itemize}

\section{A semi-analytic solution to Problem 1} 

Following the analytical results of section 3, the semi-analytic
solution to Problem 1 that we provide in this paper goes through the
following steps (we assume here that, for the integer vector $\ell$,
the function $v(\sigma)$ satisfies the hypotheses of Lemmas 1 and 2,
otherwise obvious modifications apply).
\vskip 0.4 cm
\noindent
Semi-analytic representation of $\Delta F_j$ during a resonant
  libration.  On the basis of Lemma 1 and Lemma 2 we first define the
algorithm which approximates the variation of the adiabatic actions
$\Delta F_j$ in the time interval $[0,T_\alpha]$ ($t=0$ is chosen so
that $\sigma(0)=0$ and $T_\alpha$ is the circulation period of the
dynamics of the resonant normal forms, see Section 2), with the
function (see (\ref{sumF0full}))
$$
\Delta F_j: {\Bbb T}^{n-1}\rightarrow {\Bbb R}
$$
\begin{equation}
\phi(0) \longmapsto   \sum_{m,\nu,k} f_{j,m,\nu,k}e^{ik\cdot \phi(0)}
\label{semianalyticfull}
\end{equation}
where the coefficients $f_{j,m,\nu,k}$ are provided as floating point numbers
obtained by replacing the integral in
\begin{equation}
f_{j,m,\nu,k} = -i k_j {r^m_{\nu,k}(F_*)\epsilon^{m-1\over 2}\over A}
\int_0^{2\pi}[s_\alpha(\sigma)]^{m-1}e^{i\theta(\sigma )}d\sigma
\label{fj}
\end{equation}
according to the following fast algorithm: 
\begin{itemize}
\item[-] For any $m,\nu,k$ such that $r^m_{\nu,k}(F_*)\ne 0$ 
compute $\Omega$, ${\cal N}$ and ${\cal W}$ (see (\ref{calnomega}),
(\ref{thsigpend})). 
\item[-] If ${\cal N}\cdot {\cal W}>0$, or if $|\Omega| > 1$, set
  $f_{j,m,\nu,k}=0$. If ${\cal N}\cdot {\cal W}<0$ and if $|\Omega|
  \leq 1$, check if condition (\ref{condNWalpha}) and
  $\delta<-\delta_c/2$ are satisfied. In such a case, compute
  $f_{j,m,\nu,k}$ by replacing the integral with its asymptotic
  expression as indicated in Section 3, Lemma 1.
\item[-] If $-\delta_c/2\leq \delta<\delta_c$ compute $f_{j,m,\nu,k}$
  by replacing the integral with its asymptotic expression as
  indicated in Eq.~\eqref{linearI}, with the coefficients $c_0$
    and $c_1$ given in Section 3, Lemma 2. Otherwise, if $\delta\geq
  \delta_c$ set $f_{j,m,\nu,k}=0$.
\end{itemize}

\vskip 0.4 cm
\noindent
{\bf Randomization of the phases, a refinement of equation
  (\ref{fj}).} The above representation is obtained by first
approximating the integrals in (\ref{sumF0full}) with Melnikov
integrals, and then by computing the Melnikov integrals using Lemmas 1
and 2. Then, we describe the long--term diffusion of the actions
caused by a sequence of resonant circulations by applying iteratively
formula (\ref{fj}) and by updating the values of the phases $\phi(0)$
at the beginning of each circulation, assuming a random variation. In
fact, during any resonant circulation, the angles $\phi(t)$ deviate
from the approximation considered in the Melnikov integrals,
i.e. $k\cdot \phi(t)\sim k\cdot \phi(0)+k\cdot \Omega t$. Since the
dynamics is chaotic and the phases $k\cdot \phi(t)$ are fast, we
expect a random deviation from the linear approximation $k\cdot \Omega
t$ to appear during a circulation period. The small errors introduced
by the randomization of the phases during a circulation period is
reduced if we split the period $[0,T_\alpha]$ into two time intervals:
$$
[0,\bar T_\alpha]\ \  ,\ \  [\bar T_\alpha,T_\alpha]
$$
where $\sigma(\bar T_\alpha)=\bar \sigma$, and we compute two semi-analytic
formulas: 
\begin{equation}
\sum_{m,\nu,k} f^1_{j,m,\nu,k}e^{ik\cdot \phi(0)}\ \ ,\ \ \sum_{m,\nu,k} f^2_{j,m,\nu,k}e^{ik\cdot \phi(\bar T_\alpha)}
\label{semianalytichalf}
\end{equation}
representing the change of the adiabatic actions in the first part of
the resonant circulation ($\sigma(t)\in[0,\bar\sigma]$) and in the
second part ($\sigma(t)\in[\bar\sigma,2\pi]$) respectively. The value
of the phases $\phi$ are then updated also when $\sigma=\bar\sigma$.

The reason for this improvement is the symmetry of the distribution of
the critical points with respect to the minimum $\bar \sigma$, so that
the change of the actions is really split into two well differenciated
parts, the first one taking place before $\bar\sigma$ and the second one
taking place after $\bar\sigma$.

\vskip 0.4 cm
\noindent
{\bf Largest $\Delta F_j$ during a resonant libration.} From the 
numerical values of the coefficients $f_{j,m,\nu,k},f^i_{j,m,\nu,k}$, one
can estimate the maximum variation 
$\Delta F_j$ during a resonant libration by computing 
the maximum of the function
\begin{equation}
\norm{\sum_{m,\nu,k} f_{j,m,\nu,k}e^{ik\cdot \phi(0)}}
\label{semianalyticseries}
\end{equation}
with respect to all the possible initial values of the phases $\phi(0)$.
\vskip 0.2 cm
\noindent
We remark that to obtain results which compare to the numerical
experiments we are not allowed to replace the maximum of the series
(\ref{semianalyticseries}) with the value of the majorant series
$\sum_{m,\nu,k} \norm{f_{j,m,\nu,k}}$. In fact, there are examples
(see Section \ref{numsection}) where the majorant series is one order
of magnitude larger than (\ref{semianalyticseries}).
\vskip 0.4 cm
\noindent
{\bf Orbits with the fastest long-term instability, ballistic
  diffusion.} The long-term instability of an orbit may arise from a
sequence of circulations/librations of $S,\sigma$, which produce very
small jumps of $F_j$ and $\alpha$, while the phases $\phi$ are treated
as random variables.  Since $\Delta F_1,\Delta \alpha$ are very small
at each step, their variations along several circulations/librations
are mainly determined by the values of the phases $\phi$ at the
beginning of each circulation/libration.  Random variation of the
phases yields a random walk of $F_1$ and, by selecting an initial
condition such that the values of the phases at each
circulation/libration produce the maximum $\Delta F_1$, we obtain a
monotonic ballistic motion along the resonance.  The conditions to
observe these ballistic motions from swarms of ${\cal K}$ diffusive
orbits are determined as follows: by assuming a randomization of the
phases occurring at each resonant libration (random phase
approximation), half of the orbits will have $\Delta F_1\geq 0$ and
the other half $\Delta F_1<0$, within the range determined by
$\norm{\Delta F_1(T_\alpha)}$ computed as indicated in the previous
step. Therefore, we observe orbits with $\Delta F_1$ of the same sign
for a number of ${\cal M}$ randomizations as soon as ${\cal K}/2^{\cal
  M}\geq 1$. Correspondingly, given ${\cal K}$, we observe orbits with
$F_1$ which increases (or decreases) almost monotonically in time for
a time interval $(\log {\cal K}/\log 2) T_\alpha$. By denoting with
$10^{-p}$ the precision of the numerical integration, this time
interval is bounded by $(p/\log 2)T_\alpha$.

The speed of the ballistic diffusion in the sequence of resonant librations 
is represented by
$$
{\sum_{i=1}^{\cal M}{\norm{\Delta F^{(i)}_j}}\over 
\sum_{i=1}^{\cal M}T_{\alpha^{(i)}}}  ,
$$ where we estimate the variation $\norm{\Delta F^{(i)}_j}$ occurring
at the $i$--th step by the maximum value computed using the
semi-analytic theory previously indicated. The sum of the libration
periods $T_{\alpha^{(i)}}$ is instead estimated by assuming an average
period needed by any libration
\begin{equation}\label{eq:Talph_gen}
T_{\alpha} = \int^{2\pi}_{0} \frac{d \sigma}{ \sqrt{\epsilon} \sqrt{2 |A| (M(1+\alpha) - v(\sigma))}}
\end{equation}
with $\alpha = \Norm{r^N}$. In the case $v(\sigma)=\epsilon M
\cos\sigma$, we find
\begin{equation}
T_\alpha= {1\over \sqrt{A\epsilon M}}\ln {32 A \epsilon M\over \Norm{r^N}}
\label{Talphar}
\end{equation}
obtained when the energy of the libration differs from 
the separatrix value by the norm of the remainder. 
Therefore, we obtain the formula for the average speed of the 
ballistic diffusion as
\begin{equation}
{\cal D}=\max_j {\max_{\phi(0)} \norm{\Delta F_j}\over 
{1\over \sqrt{A\epsilon M}}\ln {32 A \epsilon M\over \Norm{r^N}}}
\ \   .
\label{ballisticspeed}
\end{equation} 
\vskip 0.4 cm
\noindent
{\bf Numerical computation of $\Delta F_j(T)$ during a resonant libration.} 
The analytic formulas provided above allow us to compute the 
maximum speed of diffusion along the resonance. If we are also 
interested in following, for any given value of the phase $\phi(0)$
at the beginning of the resonant libration,  
the individual variation $\Delta F_j(T)$ for all $T\in [0,T_\alpha]$, it 
is possible to compute numerically the function
\begin{equation}
\begin{gathered}
\Delta F^N_j: [0,T_\alpha]\times  {\Bbb T}^{n-1}\rightarrow {\Bbb R} \\
 (T,\phi(0)) \longmapsto   \sum_{m,\nu,k} f_{j,m,\nu,k}(T)e^{ik\cdot \phi(0)}
\end{gathered}
\label{semianalyticnum}
\end{equation}
where the coefficients
\begin{equation}
f_{j,m,\nu,k}(T) = -i k_j {r^m_{\nu,k}(F_*)\epsilon^{m-1\over 2}\over A}
\int_0^{\sigma_0(T)}[s_\alpha(\sigma)]^{m-1}e^{i\theta(\sigma )}d\sigma
\label{fjnum}
\end{equation}
are obtained, only for the terms in the category (II) or (III), by 
evaluating numerically the integrals in (\ref{fjnum}). Since
the terms in the category (II) or (III) are just ~1/1000 of 
terms of the remainder, the numerical computation of all these
integrals is well within the possibility of modern computers.  
\vskip 0.4 cm
\noindent
    {\bf Remark:} (xvi)
    Formula (\ref{ballisticspeed}) has been obtained
from the analysis of the optimal normal form constructed as indicated
in the Nekhoroshev theorem. Therefore it represents an improvement of
the a priori estimate obtained from the same normal form, for the
diffusion along the resonances of multiplicity 1.  In the examples of
Section 5 the improvement is of some orders of magnitude.

\section{Numerical demonstrations on a three degrees of freedom steep 
Hamiltonian model}\label{numsection}

We illustrate our theory for the 3-degrees of 
freedom Hamiltonian (introduced following \cite{GLF11,TGLF})
\begin{equation}
H_\epsilon= {I_1^2\over 2}-{I_2^2\over 2}+{I_2^3\over 3\pi}+\pi I_3+{\epsilon
\over \cos\psi_1+ \cos\psi_2+ \cos\psi_3+4}~,
\label{hamexample}
\end{equation}
satisfying the hypotheses of the Nekhoroshev theorem ($H_0$ is steep
and the perturbation is analytic), for the resonance 
defined by the integer vector $\ell =(1,1,0)$, and by 
$I_*=(21\pi/100, 3\pi/10,1)$.  
\vskip 0,2 cm
\noindent
{\bf The upper value of $\epsilon$.} Using the method of the 
Fast Lyapunov Indicator (see \cite{FGL00,GL13}, FLI hereafter) 
we preliminary checked that
for the largest value of $\epsilon$ that we considered in our experiments,
i.e. $\epsilon=0.08$,  
the resonance ${\cal R}_\ell$ close to $I_*$ is embedded in a domain 
dominated by regular motions, with the other resonances forming a web,  
a circumstance ensuring that the diffusion occurs mainly 
along the resonance ${\cal R}_\ell$. Evidently, the condition persists 
for smaller values of $\epsilon$.

\vskip 0,4 cm
\noindent
{\bf Computation of the normal form using a HNA.} For a sample 
of values of $\epsilon$ we computed the normal form 
of Hamiltonian (\ref{hamexample}) by implementing the HNA 
described in \cite{efthy}. Following the notations of Lemma 2, 
for the selected resonance determined by $\ell=(1,1,0)$, 
we preliminary define the canonical transformation
$$
(\tilde S,\tilde F_1,\tilde F_{2})=\Gamma^{-T} (I_1,I_2,I_3)=(I_2,I_2-I_1,I_3)
$$
$$
(\tilde \sigma_1,\tilde \phi_1,\tilde \phi_2)=\Gamma (\varphi_1,\varphi_2,
\varphi_3)=(\varphi_1+\varphi_2,-\varphi_1,\varphi_3)
$$
with
$$
\Gamma = \left (
\begin{array}{ccc}
1 &  1& 0 \\
-1 & 0 & 0 \\
0& 0 & 1
\end{array}
\right ) ,
$$
and then, implementing the HNA, we obtain a canonical transformation
\begin{equation}
(S,F,\sigma,\phi)=\tilde {\cal C}(\tilde S,\tilde F,\tilde \sigma,\tilde 
\phi)
\label{calC}
\end{equation}
conjugating the Hamiltonian to the normal form (\ref{HN})
$$
H^N=h(F,S)+\epsilon f^N(F,S,\sigma)+r^N(F,S,\sigma,\phi)
$$ with optimal normalization order $N$ depending on the specific
value of $\epsilon$. For all the details about the HNA we refer the reader to
\cite{efthy}. Nevertheless we provide below some details about the
output of the algorithm for the case treated in this paper.
\vskip 0,4 cm 
\noindent
\begin{itemize}
\item [-]{\it Truncation order, optimal normalization order, optimal reminder.} 
Since the HNA is implemented on a computer algebra system, any
function $Z(S,F,\sigma,\phi)$ is stored in the memory of the computer as a 
Taylor--Fourier expansion defined by its series of terms\footnote{Recall
that for the specific Hamiltonian (\ref{hamexample}) the action $F_2=I_3$ 
appears only as an isolated linear term.} 
$$
(S-S_*)^{p_1}(F_1-F_{*,1})^{p_2} e^{i (\nu \sigma+k \cdot \phi)}  ,
$$
truncated to some suitably large truncation order. To define the truncation
order, as well as other orders within the algorithm, 
the series is modified by multiplying each term by
$$
\xi^{\left (p_1+p_2+{2 \mu(\norm{\nu}+\norm{k_1}+\norm{k_2})\over \ln (1/\epsilon)}\right )}
$$
where $\xi$ is a formal parameter (which at the end of the
computation will be set equal to 1), and $\mu$ 
is defined so that the perturbation is analytic 
in the complex domain $\{ \varphi:\ \norm{\Im \varphi_j}\leq \mu\}$. 
Then we represent the modified series ${\cal Z}(S,F,\sigma,\phi,\xi)$ obtained
in this way as a Taylor expansion with respect to the parameter $\xi$
$$
 {\cal Z} = \sum_{j=1}^{{\cal J}} \xi^j  {\cal Z}_j(S,F,\sigma,\phi)
$$
truncated at some suitable order ${\cal J}$. The truncation 
order of $Z$ is decided as the truncation order of the Taylor expansion 
of ${\cal Z}$ with respect to the parameter $\xi$.

The expansions in the formal parameter $\xi$ are used also to 
define the optimal normalization order. In fact, if we consider all the 
intermediate Hamiltonians which are constructed within the algorithm
$$
H^i=h(F,S)+\epsilon f^i(F,S,\sigma)+r^i(F,S,\sigma,\phi)\ \ ,\ \ i=1,\ldots ,N
$$
any remainder $r^{i}$ has a truncated Taylor expansion in the formal parameter
$$
{\cal R}^{i} = \sum_{j={\cal J}_i}^{{\cal J}} \xi^j  {\cal R}^{i}_j(S,F,\nu,\phi)
$$
starting from a minimum order ${\cal J}_i$ 
such that ${\cal J}_{i+1}={\cal J}_{i}+1$. The optimal value of $N$ 
is then chosen so that
$$
\Norm{r^1}> \Norm{r^2}>\ldots >\Norm{r^{N-1}} \geq \Norm{r^{N}}\ \ ,\ \ 
\Norm{r^{N}} < \Norm{r^{N+1}} .
$$

A necessary condition for the correct execution of the algorithm is
that the truncation order ${\cal J}$ is larger than
${\cal J}_{N}$. Therefore, the practical limitation for its implementation
is due to the limited memory of the computer to store all the series
expansions required by the HNA to work within the truncation
order. Since the optimal normalization order increases as $\epsilon$
decreases, for any given computer memory we have a lower bound on the
value of $\epsilon$ such that we are able to construct the normal form
Hamiltonian. For the practical pourpose of this work, we considered a lower
bound of $\epsilon=0.0005$.
%

\item[-] {\it Domains of the normal forms.} In order to solve
Problem 1 we need to provide an estimate of the norms of the normal form 
remainders (computed as the series of the absolute values of 
the Taylor-Fourier coefficients) 
in a domain of the actions $S,F$ which is bounded, 
in principle, by order $\sqrt{\epsilon}$. The numerical 
bounds of the action variables are chosen, for each value
of $\epsilon$, according to the amplitude of the separatrices of the 
resonant motions. 

\item[-] {\it Estimates on the canonical transformation.} The
  canonical transformation $\tilde {\cal C}$ (see (\ref{calC})) is
  near to the identity, and in particular the difference
  $\norm{F_j-\tilde F_j}$ can be uniformly bounded by $\epsilon^b$
  (with some $b>0$ defined as in (\ref{statab})) which is a quantity
  much larger than the norm of the optimal remainder $r^N$.  As a
  consequence, even if we suppress from the normal form the remainder
  $r^N$, so that the normalized actions $F_j$ are constants of
  motion, the non-normalized actions have a variation of order
  $\epsilon^b$ which cannot be ascribed to the Arnold diffusion which
  is instead produced by a variation of the normalized actions $\tilde
  F_j$ (the so--called 'deformation' in Nekhoroshev theory).
\end{itemize}
\vskip 0.4 cm

\begin{table}\label{table1}
\centering
  \begin{tabular}{|c c c c c c c|}
    \hline
    \phantom{\LARGE |} $\epsilon$ \phantom{\LARGE |} & $\Delta S$ & ${\cal J}$ &  ${\cal J}_N$ & $\Norm{r^N}$  & $T_{\alpha}$ & $|\Delta F_1|_{Nekh}$  \\
    \hline
    $\;0.08\;$ & $\;0.114\;$ & $\;9\;$ & $\;6\;$  &  $\;1.179 \snot[-4]\;$ & $\; 165.0\;$ & $ \;1.95\snot[-2]\;$\\  
    $0.05$   & $0.090$ & $9$  & $6$   &  $3.01 \snot[-5]$  & $265.9$   & $7.99 \snot[-3]$\\
    $0.02$   & $0.057$ & $10$ & $7$   &  $2.13 \snot[-6]$  & $519.6$   & $1.11 \snot[-3]$\\
    $0.01$   & $0.040$ & $12$ & $9$   &  $2.07 \snot[-7]$  & $864.2$   & $1.78 \snot[-4]$\\
    $0.008$  & $0.036$ & $13$ & $10$  &  $8.43 \snot[-8]$  & $1028.4$  & $8.67 \snot[-5]$\\
    $0.005$  & $0.029$ & $13$ & $10$  &  $1.24 \snot[-8]$  & $1468.5$  & $1.82 \snot[-5]$\\
    $0.002$  & $0.018$ & $13$ & $10$  &  $2.85 \snot[-10]$ & $2705.0$  & $7.70 \snot[-7]$\\
    $0.001$  & $0.013$ & $13$ & $10$  &  $3.05 \snot[-11]$ & $4216.4$  & $1.29 \snot[-7]$\\
    $0.0005$ & $0.009$ & $13$ & $10$  &  $4.01 \snot[-12]$ & $6442.7$  & $2.58 \snot[-8]$\\
  \hline
  \hline
 \phantom{\LARGE |} $\epsilon$ \phantom{\LARGE |} &  $|\Delta F_1|_{Max}$ & {\scriptsize (II)+(III)$_{1}$} &  $|\Delta F_1|_{P}$ & $|\Delta F_1|_{NP}$ & {\scriptsize (II)+(III)$_{2}$} & $|\Delta F_1|_{sa}$\\
  \hline
  $0.08$  & $\; 5.22\  \snot[-4] \;$  & $\; 1334 \;$  &  $\; 3.45 \snot[-4] \;$  & $\; 3.35 \snot[-4] \;$  & $\; 648 (83) \;$ & $\;5.42\snot[-4] \;$ \\
  $0.05$ & $1.75 \snot[-4]$ & $1124$ & $1.56\snot[-4]$  & $1.32\snot[-4]$ & $513(48)$   & $2.28\snot[-4]$   \\
  $0.02$ & $1.0 \snot[-5]$  & $1696$ & $1.2\snot[-5]$   & $5.2\snot[-6]$  & $601(50)$    & $8.34\snot[-5]$  \\
  $0.01$ & $2.08 \snot[-7]$  & $3838$ & $3.6\snot[-7]$   & $2.1\snot[-7]$  & $1300(202)$  & $2.91\snot[-6]$  \\
  $0.008$ & $7.0 \snot[-8]$  & $5470$ & $4.8\snot[-8]$   & $2.5\snot[-8]$  & $1703(180)$ & $4.36\snot[-7]$  \\
  $0.005$ & $1.0 \snot[-8]$  & $6476$ & $1.2\snot[-8]$   & $6.7\snot[-9]$  & $634(201)$  & $6.79\snot[-9]$  \\
  $0.002$ & $2.36 \snot[-9]$ & $7346$ & $4.38\snot[-9]$  & $2.79\snot[-9]$ & $916(7)$    & $6.66\snot[-9]$  \\ 
  $0.001$ & $1.08 \snot[-9]$ & $8350$ & $1.32\snot[-9]$  & $3.87\snot[-10]$& $885(12)$   & $7.68\snot[-10]$  \\
  $0.0005$ & $2.12 \snot[-10]$ & $9364$& $1.87\snot[-10]$ & $1.86\snot[-10]$& $836(26)$   & $2.04\snot[-10]$  \\
  \hline
  \end{tabular}
  \caption{\small Summary of the numerical experiments on Hamiltonian
  (\ref{hamexample}). The upper table reports the parameters of the
  Hamiltonian normalizing algorithm and some of the informations that
  we can extract from its output: $\Delta S$ denotes the amplitude of
  the domain in the resonant action $S$, ${\cal J}$ the truncation
  order, ${\cal J}_N$ the optimal normalization order, $\Norm{r^N}$
  the norm of the remainder expansion (\ref{remainder}) close to
  $I_*=(0.664887, 0.955495,1)$, $T_\alpha$ the period of the resonant
  variables computed using (\ref{Talphar}); $|\Delta F_1|_{Nekh}$
  represents the a {\it priori} upper bound of the maximum variation
  of $F_1$ over a period $T_\alpha$ forced by the remainder $r^N$. The
  lower table concerns the numerical computation and the analytic
  estimates about the variations of the normalized adiabatic action
  $F_1$ during a resonant period: $|\Delta F_1|_{Max}$ denotes the
  maximum variation of $F_1$ after a full resonant period for a swarm
  of 100 orbits with initial actions close to $I_*$ obtained from
  numerical integrations of the Hamilton equations; $|\Delta F_1|_{NP}$
  denotes the semi-analytic estimate of the maximum variation obtained
  by computing numerically the Melnikov integral whose phase is
  stationary or quasi-stationary (since the numerical computation of
  the Melnikov integrals is more precise than the linear
  approximation, we include for safety a larger number of terms in the
  category (III), by checking directly the value $\delta_c$ for which
  the integrals are negligible with respect to $\delta=0$; the number
  of terms, reported in the column (II)+(III)$_1$, is still in a ratio
  of $1\sim 1000$ of the total number); $|\Delta F_1|_{P}$ is
  analogous to $|\Delta F_1|_N$, but obtained with the 'patched' formula
  (\ref{patched}); $|\Delta F_1|_{sa}$ is the value obtained using the
  asymptotic expansions of Lemmas 1 or 2 ((II)+(III)$_2$ represents
  the number of terms included in this computation).  }
\end{table}

In Table 1 we summarize the values of the orders of
truncation and of optimal normalization, as well as the norm of the
optimal remainders, for a sample of values of $\epsilon$ from
$\epsilon=0.08$ down to $\epsilon=0.0005$. The computations were
performed with double floating point precision for the largest values
of $\epsilon$, and with quadruple floating point precision for the
smaller ones. The CPU time required by the execution of the HNA on a
modern fast multi-processor workstation ranges from few minutes for
$\epsilon=0.08$ to some hours for $\epsilon=0.0005$. We notice that the
norm of the optimal remainder spans 9 orders of magnitude in this
range of variation of $\epsilon$.

\begin{figure*}
\includegraphics[width=.99\textwidth]{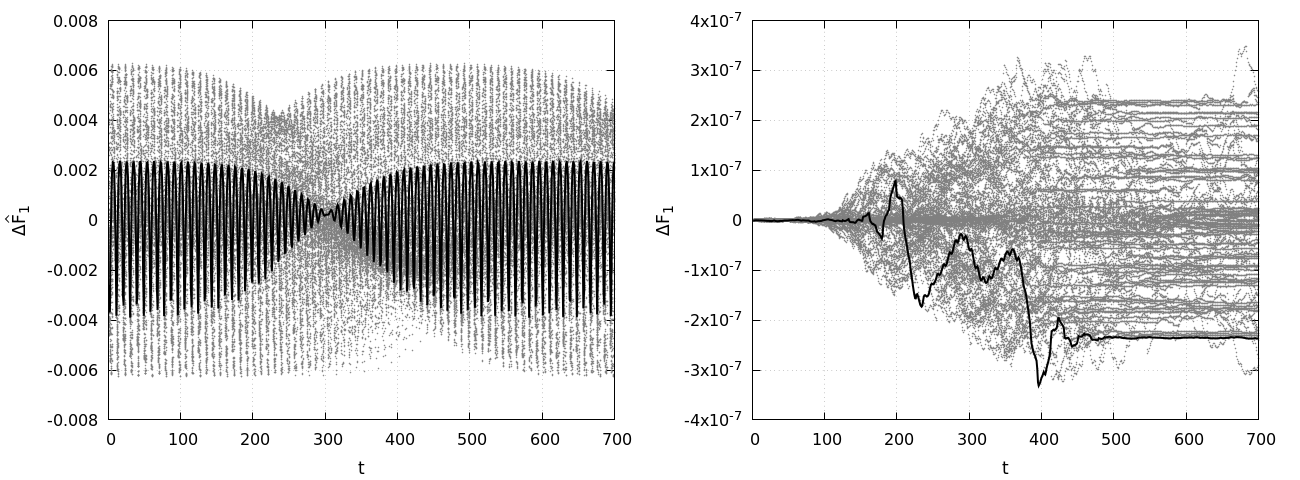}
\caption{\small Time evolution of the action $\tilde F_1$ (left panel)
  and of the normalized adiabatic action $F_1$ (right panel) for the
  same swarm of 100 solutions with initial conditions in a small
  neighborhood of the separatrix of the resonant normal form, for
  $\epsilon=0.01$. The bold curves in both panels represent the same
  sample solution.}
\label{fig:drift}
\end{figure*}

To provide an idea of the efficiency of the normalizing
transformations, in Fig.~\ref{fig:drift} we compare the time evolution
of $F_1$ with the time evolution of $\tilde F_1$ for a swarm of
solutions with initial conditions in a small neighborhood of the
separatrix of the resonant normal form, for $\epsilon=0.01$. The
solutions $(\tilde S(t),\tilde F(t),\tilde \sigma(t),\tilde \phi(t))$
have been obtained from a numerical integration of Hamilton's
equations of the original Hamiltonian (\ref{hamexample}); the
evolution of the adiabatic action $F_1(t)$ has been obtained by
transforming the numerical solution with the canonical transformation
$\tilde {\cal C}$: $(S(t), F(t), \sigma(t), \phi(t))=\tilde {\cal
  C}(\tilde S(t),\tilde F(t),\tilde \sigma(t),\tilde \phi(t))$.  In
the left panel we see that the variation of $\tilde F_1$ produces a
swarm of points rapidly oscillating in a band of width $6\times
10^{-3}$, which is due to the terms of order $\epsilon^b$ which bound
$\norm{F_1-\tilde F_1}$.  A totally different picture appears in the
right panel, where the variation of the normalized action $F_1$ is
represented: in this case the slow time evolution is well defined,
characterized by jumps of order $10^{-7}$ (typical values of
long--term diffusion of the action variables for this value of
$\epsilon$), that are detectable on such time intervals only thanks to
the implementation of the normalizing transformation.

Before applying the theory developed in Sections 2,~3,~4, we provide
as in the proof of Nekhoroshev theorem an upper bound to the variation
of the action variables by computing the right-hand side of inequality
(\ref{apriori}). The upper bound computed for a period (\ref{Talphar})
is reported in the column $\norm{\Delta F_1}_{Nekh}$ and is larger
up to two order of magnitudes with the numerically computed 
variations $\norm{\Delta F_1}_{Max}$. We therefore proceed by estimating
these variations with the Melnikov integrals.

\begin{figure*}
\includegraphics[width=.99\textwidth]{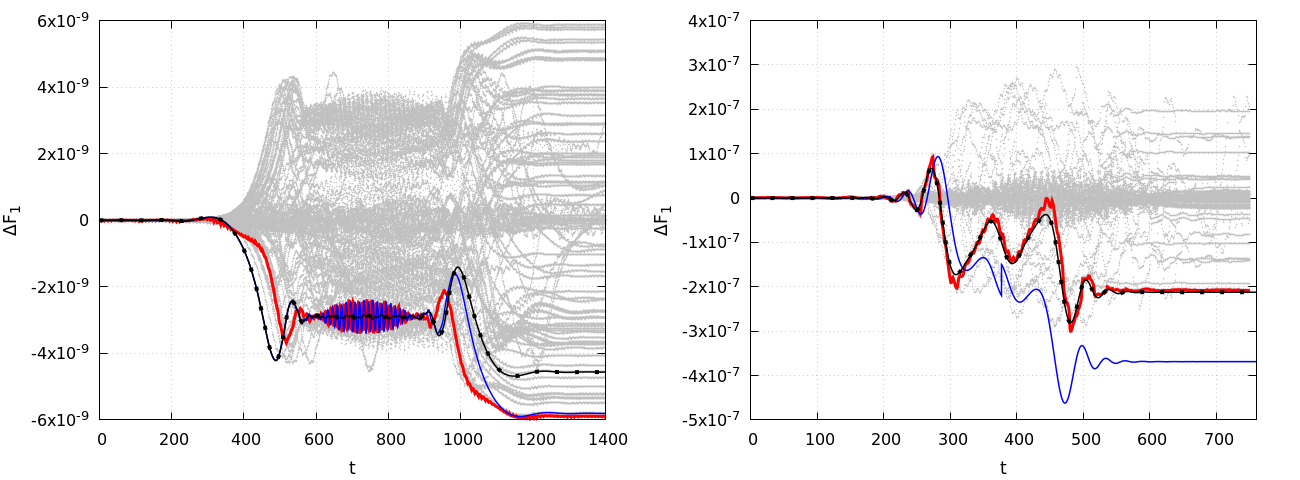}
\caption{\small Evolution of the normalized action $F_1(t)$
  numerically computed for Hamiltonian (\ref{hamexample}) with
  $\epsilon=0.003,~0.01$ (left and right panels resp.)  for a swarm of
  100 initial conditions randomly chosen in a two-dimensional square
  neighbourhood of $(S,\sigma,F,\phi)=(0,0,F_*,0,0)$ (parameterized by
  $\phi_1,S$, and with values of FLI larger than 3 over a time
  interval of $T=1000$) and performing a circulation in the $S,\sigma$
  variables. The Hamilton equations have been numerically integrated
  in the original variables $(I,\varphi)$; $F_1(t)$ has been then
  computed from the numerical solution using the canonical
  transformation defined by the HNA. The red line highlights the
  evolution with the largest $\Delta F_1$ over a circulation. The
  (dotted) black and (thin) blue lines show the evolutions obtained by
  numerically integrating the Melnikov integrals (\ref{fjnum}) whose
  phase satisfies (II) or (III), without and with the patched
  correction (\ref{patched}), respectively. }
\label{fig:drift2}
\end{figure*}

\vskip 0,4 cm
\noindent
{\bf Estimate of $\Delta F_1$ during a resonant libration.} Let us
analyze more in detail the variation of the adiabatic action $F_1$.
In Fig.~\ref{fig:drift2}, as before, we represent the time evolution
of $F_1(t)$ during a circulation of the variables $\sigma,S$ obtained
from a numerical integration of Hamilton's equations of
(\ref{hamexample}) for a swarm of 100 orbits, for two sample values of
$\epsilon$. The spread of $F_1(t)$ after the circulation is due to the
different values of $\phi(0)$. We are now able to predict the time
evolution of {\it all} these orbits by using the semi-analytic theory
developed in Section 3.

Since, due to the discrimination between phases, the number of
Melnikov integrals to take into account is now small, we have the
opportunity to compute these integrals also numerically for all the
intermediate times $t\in [0,T_\alpha]$. For these computations, we can
safely extend the value of $\delta_c$ computed from the linear
approximation as soon as the phases with $\delta>0$ provide non
negligible contributions, and still have a small number of terms (see
Table 1, column (II)+(III)$_1$).

\begin{figure*}
\includegraphics[width=.99\textwidth]{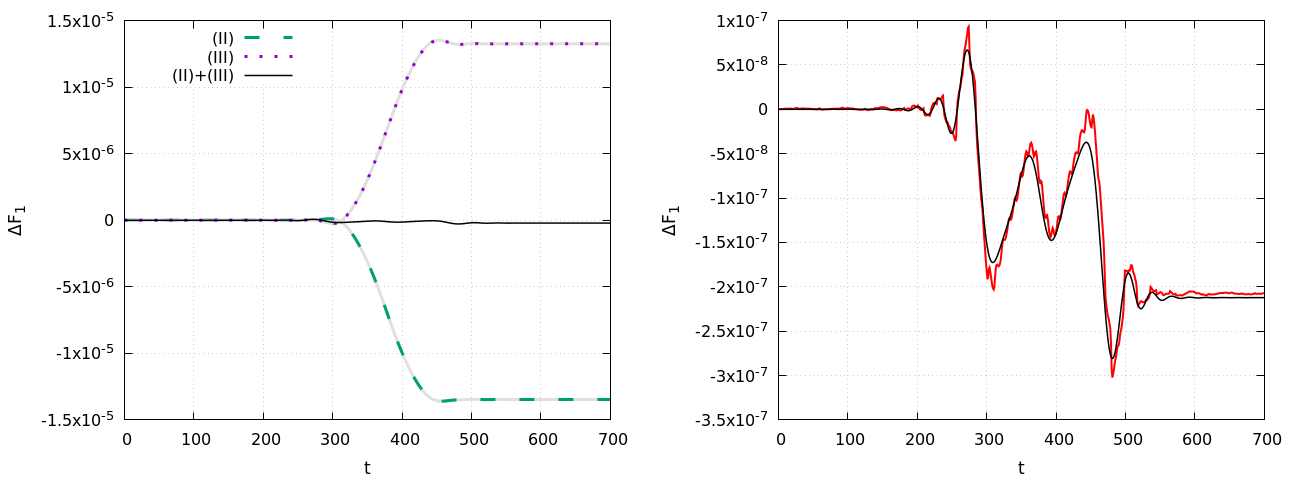}
\caption{\small Evolution of $\Delta F_1$ over a circulation of the
  resonant variables for $\epsilon=0.01$, by considering the Melnikov
  integrals whose phase is in the category (II) (dashed green line),
  in the category (III) (dotted purple line), l contribution of the
  two categories together (black line). We notice the cancellations
  occurring between the Melnikov integrals in the first and second
  case, which produce a much smaller cumulative variation, represented
  also in the zoomed right panel.}
\label{fig:cancel}
\end{figure*}

The red curves of Fig.~\ref{fig:drift2} represent the orbits yielding
the maximum negative jump obtained for the numerical integration of
the Hamilton equations, while the black and blue curves represent
the Melnikov approximations (without and with the patched formula
(\ref{patched}), respectively). One sees that for both values of
$\epsilon$ all the curves are sticked up to a time corresponding
approximately to half a period of a complete homoclinic loop. In the
middle of the homoclinic loop, we distinguish two cases. In the first
case the jump is due mostly to remainder terms which become locally
stationary at angles $\sigma_c$ sufficiently far from $\bar
\sigma=\pi$, while the slope $d\theta/d\sigma$ is substantially larger
than unity at $\sigma=\pi$. In such cases, the jumps are localized
around the two stationary values symmetric with respect to the middle
of the loop, while the associated remainder terms yield a rapid
oscillatory evolution of the actions $F_1$ in between the two jumps.
Since the motion is in reality chaotic, the orbits during the rapid
oscillations undergo also a randomization of the phases, implying that
the predictions obtained by computing (\ref{melnikint}) may introduce
an error. This can be remedied using both representations
(\ref{semianalytichalf}): precisely, the blue curve represents the
`patched' evolution given by:
\begin{eqnarray}\label{patched}
\Delta F_{1}(t)&:=&\Delta F^{N}_{1}(t)~~\mbox{if}~~t< T_\alpha/2~,\\ 
\Delta F_{1}(t)&:=&2\Delta F^N_{1}(T_\alpha/2)- \Delta F^N_{1}(T_\alpha-t)
~~\mbox{if}~~t\geq T_\alpha/2~.
\nonumber
\end{eqnarray}
On the other hand, in cases where important quasi-stationary terms
enter into play, the phase $\theta(\sigma)$ remains at small values
over a large interval around $\sigma=\pi$. Then no rapid oscillations
of the fast variables are observed, and the variations become
predictable along the whole homoclinic loop using the original
estimate (\ref{semianalyticfull}).  In fact, these are cases where the
method illustrates its full power, as it is able to capture large
cancellations taking place between stationary terms (II), which,
however, exhibit near-stationarity in the whole interval between the
two (symmetric with respect to $\pi$) critical values $\sigma_c$, and
true quasi-stationary terms (III). An example is provided in
Fig~\ref{fig:cancel}: the terms of groups (II) and (III) independently
produce jumps of order $10^{-5}$, which nearly cancel, leaving a
residual of order $10^{-7}$ which fits exactly the numerical
evolution of the action $F_1$. Since no rapid oscillations are
observed in the middle of the homoclinic loop, the non patched
estimate is more precise than the patched estimate, as also shown in
the right panel of Fig.~\ref{fig:drift2}.

In Table~1, for several different values of $\epsilon$, we report the
values $\vert\Delta F_1\vert_{Max}$ representing the maximum variation
of $F_1$ in a full resonant libration, obtained from the numerical
integration of the Hamilton equations for a swarm of 100 orbits with
initial actions close to $I_*$; $|\Delta F_1|_{NP}$ denotes the
semi-analytic estimate of the maximum variation obtained by computing
numerically the Melnikov integral whose phase is stationary or
quasi-stationary; $|\Delta F_1|_{P}$ is analogous to $|\Delta F_1|_N$,
but obtained with the patched formula (\ref{patched}); $|\Delta
F_1|_{sa}$ is the value obtained using the asymptotic expansions of
Lemmas 1 or 2. By comparing $\norm{\Delta F_1}_{Max}$ with $|\Delta
F_1|_N,|\Delta F_1|_{NP}$ we have a good agreement between the
numerical integrations and the predictive model for all the values of
$\epsilon$ (we notice that for a given $\epsilon$, only one of the two
values $|\Delta F_1|_N,|\Delta F_1|_{NP}$ is applicable), to within a
factor 2 in variations over 6 orders of magnitude as $\varepsilon$
varies between $0.0005$ and $0.08$).  The values $|\Delta F_1|_{sa}$
are expected to be slightly less precise than $|\Delta F_1|_N,|\Delta
F_1|_{NP}$, since they rely on the linear law (\ref{linearI}) for the
quasi-stationary cases, and do not take into account the patched
formula (\ref{patched}); we expect that the errors can be more
important for larger values of $\epsilon$. Here we have an agreement
within a factor 3 as $\varepsilon$ varies between 0.0005 and 0.08,
except in the interval $0.008\leq \epsilon \leq 0.02$, where the
  cancellations (as in Fig.~\ref{fig:cancel}) become important.

\vskip 0,4 cm
\noindent
{\bf Diffusion and ballistic orbits.} As discussed in Section 4, the
long-term instability of an orbit may arise from a sequence of
circulations/librations of $S,\sigma$, which produce very small jumps
of $F_1$ and $\alpha$, while the phases $\phi$ are treated as random
variables. The random variation of the phases determines the random
walk along the resonance in jumps of maximum amplitude estimated
according to the theory of Section 3; for special initial conditions
the sequence of jumps has the same sign, so that we have the orbits
which move along the resonance with the largest speed (ballistic
orbits). An illustration of this phenomenon is represented in Fig.~5
where we represent a ballistic orbit through a sequence of 14
circulations, which is the limit of the quadruple precision. The speed
of the ballistic orbits numerically measured is in agreement with
formula (\ref{ballisticspeed}) (see Table 1). Note also the
  overall random walk nature of the jumps $\Delta F_1$ for most other
  orbits nearby to the ballistic one. Since
  estimates on $\Delta F_j$ can be regarded as providing the one--step
  size in the random walk, they are crucial in modelling the diffusion
  process for a large measure of trajectories over times of practical
  interest in the applications.

\begin{figure*}
\includegraphics[width=12.5cm,height=7.5cm]{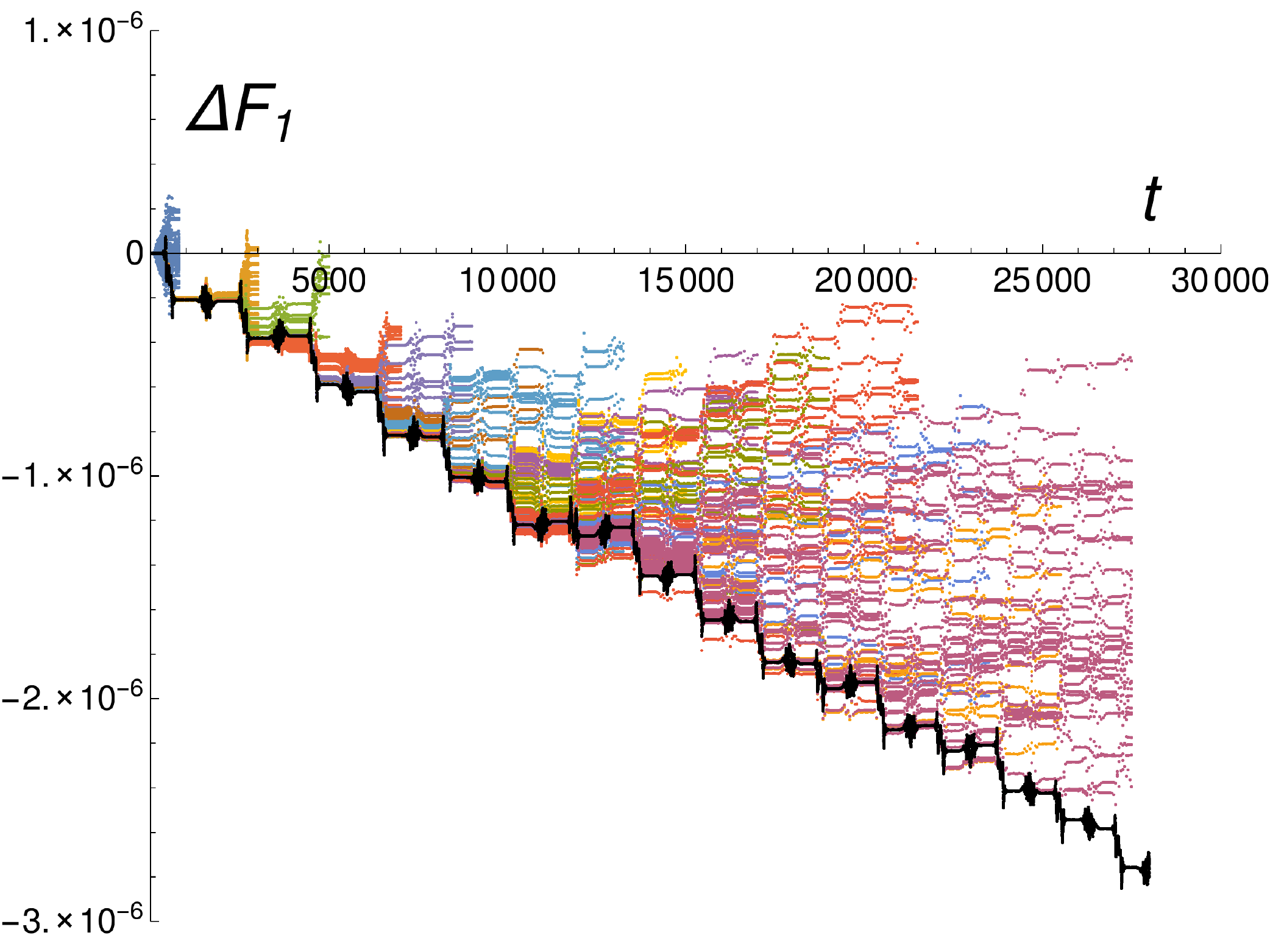}

\includegraphics[width=12.5cm,height=8.5cm]{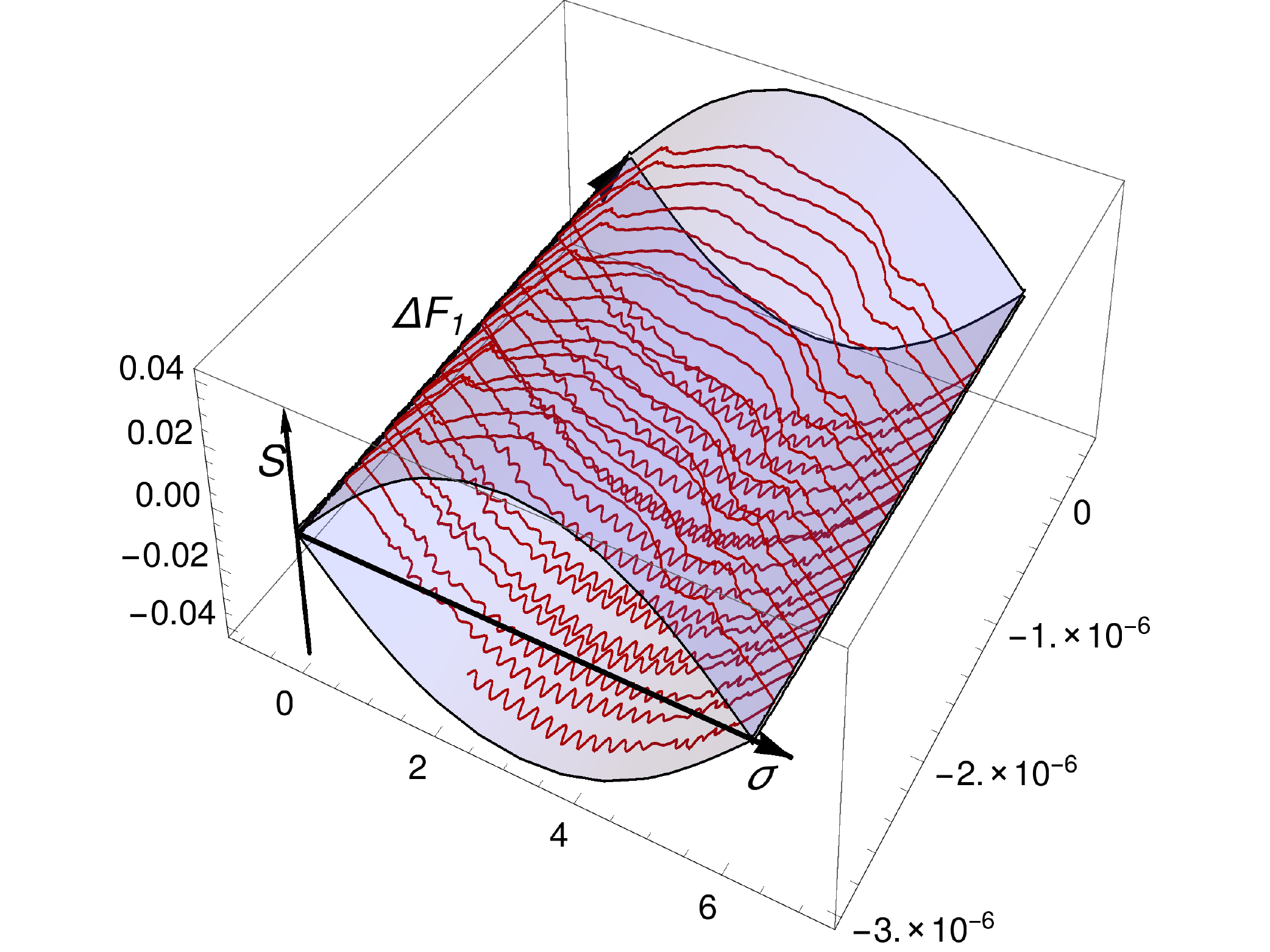}

\caption{\small Top panel: $\Delta F_1(t)$ numerically computed for
  Hamiltonian (\ref{hamexample}) with $\epsilon=0.01$ and different
  sets of ${\cal N}=16$ initial conditions in a 1-dim grid around
  $(\hat S,\sigma,F,\phi)=(0,0,F_*,\phi_*,0)$ of some selected
  amplitude $d^0$ around $\phi_*$. The initial conditions defined by
  $d^0=10^{-3}$ are numerically integrated until all the orbits
  undergo one complete circulation/libration. Then, we select
  $\phi_*^1$ in the grid providing the largest value of $\norm{\Delta
    F_1}$, and define a new set of ${\cal N}$ initial conditions in a
  1-dim grid of amplitude $d^1<<d^0$ around
  $(S,\sigma,F,\phi)=(0,0,F_*,\phi^1_*,0)$, numerically integrated
  until all the orbits complete two circulations/librations. Due to
  precision limits, the iterative refinement of initial conditions if
  stopped after 14 iterations. Different sets are reported with
  different colors, clearly illustrating the random dispersion of
  values of $F_1$, as well as the orbits with largest variations. The
  fastest (ballistic) orbit is shown in black. The bottom panel shows
  its projection in the space $S,\sigma,F_1$, thus providing a
  non-schematic example of ballistic Arnold diffusion. The diffusion
  speed is of the same order as reported in Table~1.  }
\end{figure*}

\vskip 0,4 cm
\noindent
{\bf Acknowledgments.} This research has been supported by: ERC
project 677793 Stable and Chaotic Motions in the Planetary Problem and
the Research Committee of the Academy of Athens Grant 200/854, during
non-overlapping periods (R.I.P.); project CPDA149421/14 of the
University of Padova ``New Asymptotic Aspects of Hamiltonian
Perturbation Theory" (M.G.). C.~Efthymiopoulos acknowledges the
hospitality of the University of Padova, under the program Visiting
Scientist.

\end{document}